\documentclass[a4paper,10pt]{article}

\usepackage{color}
\usepackage[usenames,dvipsnames]{xcolor} 
\definecolor{DarkGreen}{RGB}{0,64,0}

\usepackage[]{array}
\usepackage[]{amsmath}
\usepackage[]{amssymb}
\usepackage[]{mathrsfs}
\usepackage[]{tensor}
\usepackage{bm}

\newcommand{\qqd}{\ , \quad}

\newcommand{\be}{\begin{eqnarray}}
\newcommand{\ee}{\end{eqnarray}}
\newcommand{\0}{\nonumber}
\newcommand{\Gam}{\mathbf{\Gamma}}
\newcommand{\R}{\mathbf{R}}

\newcommand{\LL}{\mathbf{L}}

\newcommand{\refb}[1]{(\ref{#1})}

\newcommand{\Sent}{S}
\newcommand{\gu}{g}
\newcommand{\dg}{\delta g}
\newcommand{\gp}{\bar{g}}
\newcommand{\Hu}{H}
\newcommand{\Hp}{\bar{H}}
\newcommand{\lambdaexppar}{\alpha}

\begin{document}
\begin{flushright}
{SISSA 27/2012/EP \\ ZTF-12-02}
\end{flushright}
\vskip 1cm
\begin{center}
{\LARGE{\bf Stationary rotating black holes in theories with\\[2mm]
 gravitational Chern-Simons Lagrangian term}}
\vskip 1cm

{\Large L.~Bonora$^a$, M.~Cvitan$^b$, P.~Dominis Prester$^c$, S.~Pallua$^b$, I.~Smoli\'c$^b$}\\
{}~\\
\quad \\
{\em ~$~^{a}$International School for Advanced Studies (SISSA/ISAS),}\\
{\em Via Bonomea 265, 34136 Trieste, Italy}
{}~\\
\quad \\
{\em ~$~^{b}$Physics Department, Faculty of Science,}\\
{\em University of Zagreb, p.p.~331, HR-10002 Zagreb, Croatia}
 {}~\\
\quad \\
{\em ~$~^{c}$ Department of Physics, University of Rijeka,}\\
{\em  Ul.\ Radmile Matej\v{c}i\'{c} 2, HR 51000 Rijeka, Croatia}\\
\vskip 1cm
Email: bonora@sissa.it, mcvitan@phy.hr, pprester@phy.uniri.hr, pallua@phy.hr, ismolic@phy.hr

\end{center}

\vskip 15mm

\begin{center} {\bf Abstract} \end{center}

We study the effects of introducing purely gravitational Chern-Simons Lagrangian terms in ordinary Einstein gravity on stationary rotating black hole solutions and on the associated thermodynamical properties, in a generic number of dimensions which support these terms (i.e. in $D = 4k-1$). We analyze the conditions, namely the number of vanishing angular momenta,
under which the contributions of the Chern-Simons term to the equations of
motion and the black hole entropy vanish.  
The particular case of a 7-dimensional theory in which a purely gravitational Chern-Simons term is added to the Einstein-Hilbert Lagrangian in $D=7$ dimensions is investigated in some detail.  
As we have not been able to find exact analytic solutions in nontrivial cases, we turn to perturbation theory and calculate the first-order perturbative correction to the Myers-Perry metric in the case where all angular momenta are equal. The expansion parameter is a dimensionless combination linear in the Chern-Simons coupling constant and the angular momentum. 
Corrections to horizon and ergosurface properties, as well as black hole entropy and temperature, are presented.
\vskip 1cm 

 
\vfill\eject

\section{Introduction}
\label{sec:intro}

\medskip

Black holes are probably the most spectacular prediction of General Relativity. From a theoretical perspective, a crucial moment which lent credibility to the assumption of their existence in reality was Kerr's analytic construction of a stationary rotating black hole solution in Einstein gravity in four spacetime dimensions \cite{Kerr:1963ud}. With the development of string theory and other extra dimensions and/or higher derivative theories, it has become important to extend the Kerr solution to higher number of dimensions $D$ and/or to more general diffeomorphism covariant theories of gravity. The generalization to $D>4$, in Einstein gravity, was done by Myers and Perry in \cite{Myers:1986un}. Since then, a number of corresponding black hole solutions in different supergravity theories were constructed (for reviews see, e.g., \cite{Youm:1997hw,Maeda:2011sh}). However, despite a lot of effort, there is still not a single explicit analytic black hole solution in any generalized theory of gravity with higher curvature terms in the action in dimensions greater than three. A related problem, important also on phenomenological grounds, is that one would like to have dynamical solutions, e.g., with in-falling matter, in which a Kerr black hole is created; however so far none has been found. 

It is not hard to locate the roots for this failure of extending the Kerr solution in the abovementioned directions.  The Kerr solution (and its Myers-Perry generalization) belongs to a special class of spacetimes for which the metric can be written in Kerr-Schild form with flat seed metric. This dramatically reduces the number of unknown functions from the start. The failure of attempts that used the Kerr-Schild ansatz in some higher-curvature theories of gravity 
shows that the ansatz has limited use for black hole constructions, and that the Einstein action is somewhat special in this respect. Without some alternative simplifying property of the metric, the task of finding analytic stationary rotating black hole solutions in any $D>3$ theory seems to be hopeless. A possible strategy is to turn to different types of perturbative calculations, with the hope of extracting some information which could be useful for nonperturbative constructions.

In this paper we study asymptotically flat stationary rotating black hole solutions in theories with purely gravitational Chern-Simons terms \cite{CS} in the action in $D>3$ spacetime dimensions. One can name several reasons why these terms are interesting by themselves, including their special properties. Though they give diffeomorphism covariant contribution to the equations of motion \cite{Solodukhin:2005ns,Bonora:2011mf}, they are not manifestly diff-covariant. This leads to interesting consequences, e.g., for the black hole entropy \cite{Tachikawa:2006sz,Bonora:2011gz} and anomalies for the boundary theories (as in AdS constructions) \cite{Solodukhin:2005ns}. Topological considerations \cite{BCDPSp2} become relevant due to these terms, which moreover break parity in the purely gravitational sector. Gravitational Chern-Simons terms are present in some superstring/M theory low energy effective actions (depending on type and compactification), and though they appear more frequently in the form of mixed gauge-gravitational Chern-Simons Lagrangian terms,\footnote{The role of mixed gauge-gravitational Chern-Simons terms for black hole constructions in superstring effective theories is reviewed in
\cite{Mohaupt:2007mb,Sen:2007qy,Castro:2008ne,deWit:2009de,Prester:2010cw}. In some cases it was shown that all higher-derivative $\alpha'$-corrections to near-horizon properties of extremal black holes are originated solely by such Chern-Simons terms, though low energy effective actions contain infinite number of higher-derivative terms \cite{Prester:2008iu,Cvitan:2007hu}.} some compactifications to 7-dimensional spacetime may lead to purely gravitational Chern-Simons Lagrangian terms. It should be recalled that, despite the mentioned recent developments, there is much less understanding of the consequences of gravitational Chern-Simons terms in $D>3$, then in the simplest case of $D=3$ \cite{DJT1, DJT2} which has been thoroughly studied in the literature (for the reviews see \cite{Li:2008dq,Kraus:2006wn,Sen:2007qy}). One of the aims of this paper is to try and fill some of these gaps.

The contribution to the equations of motion due to gravitational Chern-Simons Lagrangian terms is, at least apparently, terribly involved in $D>3$. Such terms exist only in $D=4k-1$, $k \in \mathbb N$, which implies that stationary rotating black holes are characterized by $2k-1$ angular momenta. However, due to their special properties, connected to parity violation, it is possible to obtain some exact results.  For example, we show that if the solution for the metric has ``enough'' isometries (which, in the case of interest here, typically occurs when two or more angular momenta vanish) then adding a gravitational Chern-Simons term in the action does not change the black hole solutions. So, to find situations where a gravitational Chern-Simons contribution is nontrivial, one has to consider rotating black holes with at least $2k-2$ nonvanishing angular momenta. This is very complicated already in $D=7$. For this reason we have turned to perturbative calculations in a 
special case, that of a $D=7$ solution in which all angular momenta are equal. We have constructed the lowest order corrections to the Myers-Perry metric in an expansion in the Chern-Simons coupling constant and angular momentum, and we have showed that the gravitational Chern-Simons term affects all the black hole characteristics we have calculated -- horizon, ergoregion and black hole entropy (at least in this perturbative sense). Our perturbative solution does not allow expressing the metric in Kerr-Schild form with a flat seed metric. This implies that to find exact analytic solutions, if they exist, in such more general theories with gravitational Chern-Simons Lagrangian terms, one needs a new ansatz.

The outline of the paper is as follows. Section \ref{sec:eom} is devoted to establishing some general results. We show that a gravitational Chern-Simons Lagrangian term does not change stationary rotating black hole solutions and the corresponding black hole entropy if two or more angular momenta are zero. This is a consequence of the more general theorem derived in \cite{BCDPSp}. In Section \ref{sec:d7bh} we specialize to the particular theory in $D=7$ obtained by adding a gravitational Chern-Simons Lagrangian term to Einstein-Hilbert action. In Section \ref{sec:pertD7} we turn to the perturbative calculation in the Chern-Simons coupling constant, in the special case when all three angular momenta are equal. A few Appendices are devoted to details of calculations.

\vspace{10pt}

\section{A few general considerations}
\label{sec:eom}

\medskip

We are interested in gravity theories in $D=2n-1$ dimensions ($n\in2\mathbf{N}$) with 
Lagrangians of the form
\begin{equation} \label{lagrgen}
\LL = \LL_0 + \lambda \, \LL_{\mathrm{gCS}}
\end{equation}
where $\LL_0$ is some general manifestly diffeomorphism-invariant Lagrangian density
and $\LL_{\mathrm{gCS}}$ is the purely gravitational Chern-Simons (gCS) Lagrangian density given by
\begin{equation}
\label{LgCS}
\LL_{\mathrm{gCS}} = n \int_0^1 dt \; \mathrm{str} (\Gam \, \R_t^{n-1})
\end{equation}
Here $\R_t = t d\Gam + t^2 \Gam\Gam$, $\Gam$ is the Levi--Civita connection and $\mathrm{str}$ denotes a symmetrized trace, which is an example of an invariant symmetric polynomial of the Lie algebra of the $SO(1,D-1)$ group. In (\ref{lagrgen}) $\lambda$ denotes the gCS coupling constant, which is dimensionless and may be quantized \cite{BCDPSp2,Witten:2007kt,Lu:2010sj}. Since the $n=2$ ($D=3$) case is studied in detail in the literature, we shall focus on $n\ge4$ cases.

Adding gravitational CS terms to the Lagrangian brings about additional terms in the equations of motion. 
It was shown in \cite{Solodukhin:2005ns} that the equation for the metric tensor $g_{\alpha\beta}$ 
acquires an additional term $C^{\alpha\beta}$ which, for the gCS term (\ref{LgCS}), is of the form
\begin{equation}\label{gCSeom}
C^{\alpha\beta} = - \frac{n}{2^{n-1}} \ \epsilon^{\nu_1 \cdots \nu_{2n-2} (\alpha} \, \nabla_{\!\rho}  \, \left( \tensor{R}{^{\beta)}_{\sigma_1}_{\nu_1}_{\nu_2}} \, 
\tensor{R}{^{\sigma_1}_{ \sigma_2 }_{\nu_3}_{\nu_4}} \cdots 
\tensor{R}{^{\sigma_{n-3}}_{ \sigma_{n-2}}_{\nu_{2n-5}}_{\nu_{2n-4}}}
\tensor{R}{^{\sigma_{n-2}}^{\rho}_{\nu_{2n-3}}_{\nu_{2n-2}}} \right)
\end{equation}
The tensor $C^{\alpha\beta}$ is symmetric, traceless and covariantly conserved
\begin{equation} \label{Cprop}
C^{\alpha\beta} = C^{\beta\alpha} \;\;, \qquad C^\alpha_\alpha = 0 \;\;, \qquad
 \nabla_\alpha \, C^{\alpha\beta} = 0
\end{equation}
In $D=3$ $C^{\alpha\beta}$ is known as Cotton tensor, and in higher dimensions it can be regarded as some sort of generalization thereof \cite{Solodukhin:2005ns}.

The peculiar properties of gCS terms make them rather special. They have 
a topological character (leading to quantization of their coupling constant), they are not manifestly diffeomorphism covariant but their contribution to equations of motion (\ref{gCSeom}) is diff-covariant, they are parity-odd, and conformally covariant \cite{CS,Solodukhin:2005ns}. We are interested in investigating how they affect black hole solutions found in theories where they are absent, once they are added to the theory. However, as we elaborated in \cite{BCDPSp,Bonora:2011mf}, it appears that 
it is not easy to find physically interesting configurations for which the gCS contribution to the equations of motion (\ref{gCSeom}) is nonvanishing and are at the same time simple enough to be analytically tractable.\footnote{Notable exceptions are nontrivial analytically tractable solutions obtained in \cite{Lu:2010sj} by ``squashing'' maximally symmetric spaces. Such solutions may play a role in AdS/CFT constructions.} In \cite{BCDPSp} we proved a theorem  for any metric in $D$ dimensions of the form
\begin{equation} \label{vaneom}
ds^2 = g_{\mu\nu}(x) \, dx^\mu dx^\nu = g_{ab}(y) \, dy^a dy^b + f(y) \, h_{ij}(z) \, dz^i dz^j 
\end{equation}
where local coordinates are split as $x^\mu = (y^a, z^i)$, $\mu = 1,\ldots,D$, $a=1,\ldots,d$,
and $i=1,\ldots,p$ ($d+p=D$), and $g_{ab}(y)$ and $h_{ij}(z)$ are arbitrary tensors depending only on the
$\{y^a\}$ and $\{z^i\}$ coordinates, respectively. It turns out that if $d>1$ and $p>1$ the gCS contribution to the equations of motion vanishes, i.e.,
\begin{equation}
C^{\mu\nu}[g] = 0 \;\;. 
\end{equation}
Due to the conformal covariance of the $C^{\mu\nu}$ tensor, the theorem extends to any metric which is conformally equivalent to (\ref{vaneom}).

As discussed in \cite{BCDPSp}, this theorem covers many classes of metrics usually discussed in the
literature. In particular, it also applies to all spacetimes with local $SO(k)$ isometry, with $k\ge 3$. It 
appears that if we want to study gCS Lagrangian terms with nontrivial influence, stationary rotating asymptotically flat black hole solutions are the next simplest objects.

Introducing additional terms in the action generally affects also asymptotic charges, such as mass $M$ and angular momenta $J_i$. For our purposes the most convenient method appears to be the one based on the energy-momentum pseudotensor (see, e.g., section 7.6 of \cite{Weinberg} or \cite{Deser:2002jk}) in which $M$ and $J_i$ are obtained by integrating over a $(D-2)$ surface $\mathcal{S}_\infty$ (the asymptotic spacelike boundary) some linear functional of a deviation of the metric from the background metric, which is obtained from the linearized equations of motion. In this paper we are primarily interested in the case of asymptotically flat metrics, where the background metric is the Minkowski one.\footnote{All the metrics we consider in this paper have ``standard'' asymptotic behavior, which makes us confident in using the energy-momentum pseudotensor method. As an independent check, we have shown that the first law of black hole thermodynamics is satisfied in all the cases where the calculation is possible (i.e., in Sections \ref{sssec:two0a} and \ref{sssec:1stlaw}). We thank the referee of this paper for proposing this check.} In this case it is obvious from (\ref{gCSeom}) that the gCS Lagrangian term with $n > 2$ (i.e., in $D>3$) does not contribute to the linearized equations of motion, which means that the formal expressions for asymptotic charges are the same as in the theory without gCS Lagrangian term. In mathematical terms
\begin{equation} \label{Qpseudo}
Q[h] = Q_{0}[h] = Q_{0}[h_0] + Q_{0}[h_{\mathrm{gCS}}] \;, \qquad 
g_{\mu\nu} = \eta_{\mu\nu} + h_{(0)\mu\nu} + (h_{\mathrm{gCS}})_{\mu\nu}
\end{equation}
The only possible effect of gCS Lagrangian term on $M$ and $J_i$ is indirect and affects the solution for the metric $g_{\mu\nu}$ through its contribution to the equations of motion.\footnote{This is not true in three dimensions, because the gCS Lagrangian term with $n=2$ affects the linearized equations of motion.} We see that in the special case when $(h_{\mathrm{gCS}})_{\mu\nu} = 0$, i.e., when the gCS term does not affect the solution for the metric, mass and angular momenta are also unaffected
\begin{equation}
M = M_0 \;\;, \qquad J_i = J_{(0)i}
\end{equation}

We shall be interested also in thermodynamics of black holes. It was shown in 
\cite{Tachikawa:2006sz,Bonora:2011gz} that a gCS Lagrangian term (\ref{LgCS}) brings in an additional term in the black hole entropy formula. For a theory with Lagrangian (\ref{lagrgen}) the latter is given by
\begin{equation} \label{entgen}
S = S_0 + \lambda\, S_{\mathrm{gCS}} \;\;.
\end{equation}
$S_0$ is Wald black hole entropy \cite{Iyer:1994ys} due to the Lagrangian $\LL_0$. In coordinate systems of the type standardly used in the literature (like the generalized Boyer-Lindquist type of coordinates we 
use in this paper) $S_{\mathrm{gCS}}$ can be calculated from
\begin{equation} \label{entgCS}
S_{\mathrm{gCS}}[g] =  4\pi n \int_\mathcal{B} \Gam_N \R_N^{n-2} \;\;,
\end{equation}
where $\mathcal{B}$ is the $(D-2)$-dimensional bifurcation surface of the black hole horizon and 1-form
$\Gam_N$ and 2-form $\R_N$ are defined in Appendix \ref{app:tech:S1g0} \cite{Bonora:2011gz}. 
In a forthcoming paper, \cite{BCDPSp}, by using conformal invariance of (\ref{entgCS}), we shall prove a theorem according to which, for black hole metrics of the form (\ref{vaneom}), with $p \ge 1$ and coordinates $z$ tangential to the bifurcation surface of the horizon, the gCS entropy term (\ref{entgCS}) vanishes.

Using the just mentioned theorems, we can already state one general result. If for stationary rotating
black hole $p$ of angular momenta $J_i$ are zero, then the spacetime usually has $SO(2p)$ isometry.
Let us restrict to the cases in which this is valid.\footnote{We restrict ourselves here to ``standard'' black holes with horizon topology given by a sphere $S^{D-2}$. In this case the above symmetry statement is valid when there is no matter outside the horizon. However, it can be violated if there is matter with symmetry breaking energy-momentum tensor (e.g., rigid matter which does not rotate in corresponding directions but with the shape which breaks the $SO(2p)$ isometry). Such systems are excluded in our analysis.} Then, if $p\ge2$ such spacetime falls under the class of the above theorems guaranteeing 
$C^{\mu\nu} = 0$ and $S_{\mathrm{gCS}} = 0$. This leads us to the following clearcut statement:

\medskip

\noindent
\emph{If in the theory with some arbitrary Lagrangian $\LL_0$, a solution has two or more vanishing angular momenta $J_i$, then introducing a Lagrangian gCS term (as in (\ref{lagrgen})) does not change the solution nor the corresponding black hole entropy. Moreover, if the metric is asymptotically flat, then mass and angular momenta of the configuration also remain unchanged.}\footnote{In this case, if the solution is a black hole all thermodynamical parameters and potentials are unaffected by gCS Lagrangian term.}

\medskip

If the black hole solution with only one vanishing angular momentum  is also of the form 
(\ref{vaneom}), then by the second theorem the gCS entropy term (\ref{entgCS}) again vanishes. However, 
though this indeed applies to \emph{all known} stationary rotating black hole solutions (e.g., the Myers-Perry black holes we discuss in the next section), for the general Lagrangian (\ref{lagrgen}) there is no guarantee that solutions with only one angular momentum vanishing are of the form (\ref{vaneom}). Indeed, we shall show in the next section on an explicit example that, when only one angular momentum is vanishing, a gCS term, due to its parity-odd structure, forces the solution to depart from the form (\ref{vaneom}).

In conclusion, we see that if we want to study the problem in which gCS Lagrangian terms have non-trivial influence on stationary rotating black hole solutions, we cannot take more then one angular momentum to be zero, because in those cases both solution and entropy are unchanged when we ``switch on" coupling 
constant $\lambda$ in (\ref{lagrgen}). If only one angular momentum is zero, the solution is generally affected, but the first order correction in gCS coupling $\lambda$ of the gCS entropy term vanishes. So, to find a \emph{completely} non-trivial problem, in which all interesting ingredients are non-vanishing, we need to analyze black holes with all angular momenta nonvanishing. If we add to this that in $D=3$ dimensions it is known that a gCS term does not change rotating black hole solutions such as BTZ black hole (though it contributes to horizon and asymptotic charges such as entropy, mass and angular momentum), it follows that we have to go to $D \ge 7$ dimensions.

\vspace{10pt}

\section{Stationary rotating black holes in $D=7$}
\label{sec:d7bh}

\medskip

Following the conclusion of the previous section, from now on we specialize to the simplest non-trivial 
case with action 
\begin{equation} \label{lhecs}
\LL = \LL_0 + \lambda \, \LL_{\mathrm{gCS}}
 = \frac{1}{16 \pi G_N} \boldsymbol{\epsilon} R + \lambda \, \LL_{\mathrm{gCS}}.
\end{equation}
Such theory in $D=3$ is known as topologically massive 
gravity and was first considered in \cite{DJT1, DJT2}. We are interested in finding stationary rotating asymptotically flat black hole solutions in $D=7$.

\vspace{5pt}

\subsection{Myers-Perry black holes}
\label{ssec:MPBH}

For $\lambda=0$ we have ordinary general relativity with Einstein-Hilbert Lagrangian for which 
stationary rotating asymptotically flat black holes, with the horizon topology of the 5-sphere $S^5$, are described by Myers-Perry solutions (MP BH) \cite{Myers:1986un,Myers:2011yc}. Here we review the basic properties of Myers-Perry solutions we shall need in our calculations. 

In generalized Boyer-Lindquist coordinates the MP metric in $D=7$ is given by
\begin{eqnarray} \label{mpbh}
ds_{\mathrm{MP}}^2 = -dt^2
 + \frac{\mu \, r^2}{\Pi \, F} \left( dt - \sum_{i=1}^3 a_i \mu_i^2 \, d\phi_i \right)^{\!2}
 + \frac{\Pi \, F}{\Pi - \mu \, r^2} dr^2 + \sum_{i=1}^3 (r^2 + a_i^2) (d\mu_i^2 + \mu_i^2 d\phi_i^2)
\end{eqnarray}
where
\begin{equation} \label{fpidef}
F = F(r,\vec{\mu}) = 1 - \sum_{i=1}^3 \frac{a_i^2 \, \mu_i^2}{r^2 + a_i^2} \; , \qquad \qquad
\Pi = \Pi(r) = \prod_{i=1}^3 (r^2 + a_i^2)
\end{equation}
and the coordinates $\mu_i$ are not all independent but satisfy
\begin{equation} \label{mucon}
\sum_{i=1}^3 \mu_i^2 = 1 \;\;.
\end{equation}
From the asymptotic behavior of the metric (\ref{mpbh}) it can be shown \cite{Myers:1986un} that 
four free parameters $\mu$ and $a_i$ ($i=1,2,3$) determine the mass $M$ and angular momenta 
$J_i$ with
\begin{eqnarray}
M &=& \frac{5 \, \pi^2}{16 \, G_N} \mu \label{MPmass} \;\;, \\
J_i &=& \frac{\pi^2}{8 \, G_N} \mu \, a_i = \frac{2}{5} M a_i  \label{MPangm} \;\;.
\end{eqnarray}
We shall assume $\mu > 0$ from now on. The event horizon of the MP BH is located at $r = r_H$ where the horizon radius $r_H$ is the largest solution of the polynomial equation
\begin{equation} \label{rh0eq}
\Pi(r_H) - \mu \, r_H^2 = 0 \;\;.
\end{equation}
Eq. (\ref{rh0eq}) is a cubic equation in $r^2$, with three solutions which we denote 
$r_{\mathrm{min}}^2$, $r_-^2$, and $r_{\mathrm{max}}^2 \equiv r_H^2$. The exact expressions for
roots is rather awkward (see \cite{Doukas:2010be}) and we shall not use it. For later purposes we
note the obvious relation (obtained from one of Vieta's formulae)
\begin{equation} \label{rootsprod}
r_{\mathrm{min}}^2 \,r_-^2 \, r_H^2 = - (a_1 a_2 a_3)^2 \;\;.
\end{equation} 
To keep our analysis simple we restrict to the case in which the largest solution satisfies
$r_{\mathrm{max}}^2 = r_H^2 > 0$.\footnote{For a discussion of the subtleties of extending spacetime 
to the $r^2 < 0$ region see \cite{Myers:1986un} and a review \cite{Myers:2011yc}.} A necessary, but not sufficient, condition for this is $\mu > \sum_i \prod_{j \ne i} a_j^2$. In this case all the roots are real, and satisfy $r_{\mathrm{min}}^2 < 0 \le r_- \le r_H^2$. The surface defined by $r=r_-$ is the inner horizon, which is hidden from the outside observer by event horizon $r = r_H$.

Using (\ref{rh0eq}) and (\ref{mpbh}) one obtains that the horizon area is given by 
\begin{equation}
A_H = \pi^3 \mu \, r_H \;\;.
\end{equation} 

The ergosurface is an infinite redshifted surface, located outside the event horizon, defined by the
condition $g_{tt} = 0$, which for MP BH metric (\ref{mpbh}) leads to an equation
\begin{equation} \label{MPergo}
\Pi(r) \, F(r,\vec{\mu}) = \mu \, r^2 \;\;.
\end{equation}

As we are interested in black hole thermodynamics, let us quote that the entropy $S$, temperature 
$T$, and angular velocities $\Omega_i$ of the MP BH are given by
\begin{eqnarray} \label{MPent}
S &=& \frac{A_H}{4 \, G_N} = \frac{\pi^3}{4 \, G_N} \mu \, r_H \;\;,  \\
T &=&  \frac{\kappa}{2\pi} = \frac{\Pi'(r_H) - 2\mu \, r_H}{4\pi \mu \, r_H^2}  \;\;, 
\label{MPtemp} \\
\Omega_i &=& \frac{a_i}{r_H^2 + a_i^2} \;\;.
\label{MPomega}
\end{eqnarray}
MP black holes with coincident inner and outer horizon radii, $r_- = r_H$, obviously have $T=0$, which means that they are extremal black holes.

A general MP BH in $D=2m+1$ with generic choice of parameters $\mu$ and $\vec{a}$ is quite 
complicated to analyze. One reason is that for generic choice of the parameters $\mu$ and $\vec{a}$ one 
has a rather ``modest'' isometry group $\R \times U(1)^m$. There are two mechanisms by which one can straightforwardly enlarge the isometry group in a simple way and/or simplify calculations: 
\begin{itemize}
\item[(a)]
Taking $k$ of the angular momenta $J_i$ vanishing, which for a MP black hole means taking the corresponding $a_i$ to vanish. This enlarges the factor $U(1)^k$ to $SO(2k)$ in the isometry group. 
\item[(b)] 
Taking $k$ of the angular momenta $J_i$ to be equal, which for a MP black hole means taking the corresponding $a_i$ to be equal. This enhances the factor $U(1)^k$ to $U(k)$.
If all $J_i$ are equal, then we obtain cohomogeneity-1 metrics in which all ``angular" dependence is determined, and the only freedom left is in a number of functions of the radial coordinate $r$.
\end{itemize}

In case (a), already if just one $a_j = 0$, a direct consequence is that the radius of the inner 
horizon is $r_- = 0$, and the polynomial in (\ref{rh0eq}) is of one order smaller, which simplifies solving for the event horizon radius $r_H$. In the case of our main interest, $D=7$, by taking $a_3 = 0$ we obtain
\begin{equation}
r_H = \frac{1}{\sqrt{2}}\left( -(a_1^2 + a_2^2) + \sqrt{4 \mu + (a_1^2 - a_2^2)^2} \right)^{1/2}
\end{equation}
where a (necessary and sufficient) condition to have $r_H^2 > 0$ is $\mu > a_1^2 a_2^2$.
We can now make further simplifications either by applying (a) again, or (b). By taking also $a_2 = 0$ 
the isometry group is enlarged from $\R \times U(1)^3$ to $\R \times U(1) \times SO(4)$. If, on the other hand, we restrict to $a_1 = a_2 \equiv a$, then the symmetry is enlarged to $\R \times U(1) \times U(2)$ and we obtain a simple expression for $r_H$
\begin{equation}
r_H = \left( \sqrt{\mu} - a^2 \right)^{1/2}
\end{equation}

Another variant of the possibility (c) in $D=7$ is to have all three parameters $a_i$ equal, 
$a_1 = a_2 = a_3 \equiv a$, with isometry group $\R \times U(3)$. From (\ref{rh0eq}) and 
$\mu > 0$ then it follows that $r_H^2 > 0$ requires $\mu > 27\, a^4/4$. From (\ref{fpidef}) and (\ref{mucon}) it follows that $F$ is a function of $r$ only
\begin{equation} \label{Faequal}
F = F(r) = 1 - \frac{a^2}{r^2 + a^2} = \frac{r^2}{r^2 + a^2}
\end{equation}
which, together with (\ref{MPergo}), yields an especially simple expression for the location of the ergosurface: $r = r_e$, where
\begin{equation}
r_e = \left( \sqrt{\mu} - a^2 \right)^{1/2}
\end{equation}

\vspace{5pt}

\subsection{Adding gCS Lagrangian terms}
\label{ssec:gCSexact}

We now turn our attention to the full Lagrangian (\ref{lhecs}) with $\lambda \ne 0$, for which we would like to find solutions describing stationary rotating black holes which we denote $\bar{g}_{\mu\nu}$. Equations
of motion now read
\begin{equation} \label{eomD7l}
R^{\alpha\beta} - \frac{1}{2} g^{\alpha\beta} R - 16\pi \, G_N \lambda \, C^{\alpha\beta} = 0
\end{equation}
where $C_{\mu\nu}$ is the contribution of the gCS term which in $D=7$ is obtained by putting 
$n=4$ in (\ref{gCSeom})
\begin{equation} \label{cD7}
C^{\alpha\beta} = - \frac{1}{2} \ \epsilon^{\nu_1 \cdots \nu_6 (\alpha} \, \nabla_{\!\rho}  \, 
\left( \tensor{R}{^{\beta)}_{\sigma_1}_{\nu_1}_{\nu_2}} \, 
\tensor{R}{^{\sigma_1}_{ \sigma_2 }_{\nu_3}_{\nu_4}} \, 
\tensor{R}{^{\sigma_2}^{\rho}_{\nu_5}_{\nu_6}} \right)
\end{equation}
Contracting (\ref{eomD7l}) with $g_{\alpha\beta}$ and using the fact that $C^{\alpha\beta}$ is traceless, 
(\ref{Cprop}), it follows that $R = 0$. Inserting this back in (\ref{eomD7l}) we obtain the equations of motion in simpler form
\begin{equation} \label{eomD7}
R^{\alpha\beta} - 16\pi \, G_N \lambda \, C^{\alpha\beta} = 0 \;\;.
\end{equation}

The entropy is given by
\begin{equation} \label{entD7}
S[\bar{g}] = S_0[\bar{g}] + \lambda S_{\mathrm{gCS}}[\bar{g}]
 = \frac{A_H[\bar{g}]}{4 \, G_N} + 16\pi \lambda \int_\mathcal{B} \Gam_N[\bar{g}] \R_N[\bar{g}]^2 \;\;,
\end{equation}
where $A_H[\bar{g}]$ is the horizon area calculated from the metric $\bar{g}_{\mu\nu}$ which is a 
solution to the full equations of motion (\ref{eomD7}). It is convenient for later discussions to write solutions of (\ref{eomD7}) in the following form
\begin{equation}
\bar{g}_{\alpha\beta} = g_{(0)\alpha\beta} + \delta g_{\alpha\beta} \;\;, \qquad \quad
 g_{(0)\alpha\beta} = (g_{\mathrm{MP}})_{\alpha\beta}
\end{equation}
where $g_{\mathrm{MP}}$ is Myers-Perry black hole, which is a solution for $\lambda = 0$. For a
generic MP black hole metric we obtain (see Appendix \ref{app:tech:S1g0})
\begin{equation} \label{entgCSMP}
S_{\mathrm{gCS}}[g_{\mathrm{MP}}] = 128 \, \pi^4 \frac{\mu}{r_H} a_1 a_2 a_3
 \left( \sum_{i=1}^3 \frac{1}{r_H^2 + a_i^2} \right)^{\!3}
\end{equation}
Observe that (\ref{entgCSMP}) automatically vanishes when one or more angular momentum 
parameters $a_i$ vanish. The result (\ref{entgCSMP}) is especially interesting when 
$\delta g_{\alpha\beta} = 0$, in which case it gives the full gCS contribution to the black hole entropy. In generic 
cases, when $\delta g_{\alpha\beta} \ne 0$, it gives a part of the first-order correction to the black hole entropy in the perturbative expansion in $\lambda$ (the second part comes from $S_0[\bar{g}]$ term in (\ref{entD7}).)

There is little hope to find exact solutions with generic angular momenta of such highly involved field equation as (\ref{eomD7})-(\ref{cD7}). There are some conclusions that can be generalized from the perturbative analysis of the special case $J_i = J$, $i=1,2,3$, presented in Section \ref{sec:pertD7}.  The gCS Lagrangian term generically changes the metric and all the geometric and thermodynamic parameters (the exceptions are commented below), aside possibly from the mass $M$ and the angular momenta $J_i$. We show that for black holes with $J_i = J$, $M$ and $J$ are still given by the MP expressions 
(\ref{MPmass})-(\ref{MPangm}) up to first-order in the gCS coupling $\lambda$. This leads us to the conjecture that this is true to all orders in $\lambda$ for all the black holes we study here.

We now turn to analysis of special cases with enhanced isometry group, and thereafter we turn to perturbative calculations.

\subsubsection{$a_2 = a_3 = 0$, $a_1 \ne 0$}  
\label{sssec:two0a}

Let us us start with the most symmetric case involving rotating black holes in $D=7$. As noted in 
Sec. \ref{ssec:MPBH}, when two angular parameters are zero, e.g., $a_2 = a_3 = 0$, the symmetry of the MP metric is enhanced to $\R \times U(1) \times SO(4)$. From the general discussion in Sec. \ref{sec:eom} (see statement on page 5) we 
then know that the solution, its mass and angular momenta, and all the thermodynamical parameters including 
the black hole entropy remain the same as in the $\lambda = 0$ case. This means
\begin{equation} \label{gtwo0a}
\bar{g}_{\alpha\beta} = g_{(0)\alpha\beta} = (g_{\mathrm{MP}})_{\alpha\beta}
\end{equation}
and mass, angular momenta, entropy, temperature and angular velocities are obtained by putting 
$a_2 = a_3 = 0$ into (\ref{MPmass}), (\ref{MPangm}), (\ref{MPent}), (\ref{MPtemp}), (\ref{MPomega}), respectively. In particular, one gets that two angular momenta ($J_2$ and $J_3$) vanish, while the black hole entropy is
\begin{equation} \label{enttwo0a}
S[\bar{g}]  = S_0[g_{\mathrm{MP}}] 
= \frac{\pi^3 \mu}{4\sqrt{2} \, G_N} \, \left( \sqrt{4 \mu + a_1^4} - a_1^2 \right)^{1/2} \;\;.
\end{equation}
where $S_0$ is the Bekenstein-Hawking entropy and $g_{\mathrm{MP}}$ is the MP black hole with $a_2 = a_3 = 0$. As a check, we see that the result (\ref{entgCSMP}) in this case gives
\begin{equation} \label{entCStwo0a}
S_{\mathrm{gCS}}[\bar{g}]  = S_{\mathrm{gCS}}[g_{\mathrm{MP}}] = 0
\end{equation}
which is consistent with (\ref{enttwo0a}). If we want to see nontrivial effects of the gCS Lagrangian term we have to go to less symmetric cases.

\subsubsection{$a_3 = 0$,  $a_1 \ne 0$, $a_2 \ne 0$}  
\label{sssec:one0a}

Now we take just one vanishing angular parameter, e.g., $a_3$ (so $a_3 = 0$,  $a_1 \ne 0$, 
$a_2 \ne 0$). In this case in general there is no important enhancement of the symmetry group of isometries of MP metric. For the corresponding MP black hole by explicit calculation we have established that
\begin{equation} \label{cmnnv}
C^{\alpha\beta}[g_{\mathrm{MP}}] \ne 0 \;\;, \qquad 
 \textrm{when \ $a_3 = 0$ ,  $a_1 \ne 0$ ,  $a_2 \ne 0$}
\end{equation}
so a gCS contribution to the equations of motion are in this case nontrivial and MP black holes are no longer solutions, i.e.,
\begin{equation}
\bar{g}_{\alpha\beta} \ne (g_{\mathrm{MP}})_{\alpha\beta} \;\;, \qquad 
 \textrm{when \ $a_3 = 0$ ,  $a_1 \ne 0$ ,  $a_2 \ne 0$} 
\end{equation}
The equations of motion still look too complicated to offer much hope for finding exact solutions. However, we can get some information from a perturbative analysis. Direct calculation shows that nonvanishing 
components in (\ref{cmnnv}) are $C^{t\phi_3}[g_{\mathrm{MP}}]$, $C^{\phi_1\phi_3}[g_{\mathrm{MP}}]$
and $C^{\phi_2\phi_3}[g_{\mathrm{MP}}]$, which shows that a perturbative solution (around $\lambda = 0$) is not of the form (\ref{vaneom}) when $\lambda \ne 0$. 

Let us turn our attention to the black hole entropy. If we plug the MP metric with $a_3 =0$ into the gCS entropy term, from (\ref{entgCSMP}) we obtain
\begin{equation} \label{entMP1a0}
S_{\mathrm{gCS}}[g_{\mathrm{MP}}] = 0 \;\;, \qquad \textrm{(for a MP BH with } a_3 = 0)
\end{equation}
It follows that up to first-order in a perturbative expansion in $\lambda$, the black hole entropy is given by
Bekenstein-Hawking area formula. However, as a perturbed solution is not of the form (\ref{vaneom}), it
is possible that a gCS entropy term gives nonvanishing contribution starting from second order in 
$\lambda$.

\subsubsection{$a_i = a \ne 0$ for all $i=1,2,3$}  
\label{sssec:allJequal}

The case in which all angular momenta are equal and nonvanishing deserves a special place. On the one 
hand, it keeps all the non-trivial consequences of the most generic case. This means that all quantities (except charges defined at asymptotic infinity), both geometric and thermodynamic, are affected by the presence of the gCS Lagrangian term.\footnote{We shall show this explicitly in Sec. \ref{sec:pertD7}.} 
On the other hand the symmetry group of isometries enhances to $\R \times U(3)$
which induces significant constraints on the metric. This combination makes this case an ideal laboratory for calculations, and we shall explore it in detail perturbatively in Sec. \ref{sec:pertD7}. 

We have already shown how results in this case simplify for $\lambda = 0$, which is for MP black holes 
with $a_i = a \ne 0$, $i=1,2,3$.  Let us just note that the result (\ref{entgCSMP}) also simplifies and becomes
\begin{equation} \label{sCSa0}
S_{\mathrm{gCS}}[g_{\mathrm{MP}}] = 3456 \, \pi^4 \left( \frac{a}{r_H} \right)^{\!3}
\end{equation}
where $r_H$ is horizon radius of MP black hole.

\medskip

\noindent
In Table \ref{table:1} we summarize our results in a compact form.

\begin{table}
\label{table:1}
\begin{tabular}{cccc}
\hline
rotational parameters & MP solves EOM & entropy ($S$) & $M$ and $J_i$ \\
\hline
$a_2 = a_3 = 0, a_1 \neq 0$ & yes, see \refb{gtwo0a} & $A/4$ (MP, exact) & MP (exact) \\
$a_3 = 0, a_1 \neq 0, a_2 \neq 0$ & no & $\bar{A}/4$ & ? \\
$a_i = a \neq 0$ & no & $\bar{A}/4 + \lambda  
S_{\mathrm{gCS}}[g_{\mathrm{MP}}]$, see \refb{sCSa0} & MP \\
general $a_i$ & no & $\bar{A}/4 + \lambda  
S_{\mathrm{gCS}}[g_{\mathrm{MP}}]$, see \refb{entgCSMP} & ? \\
\hline
\end{tabular}
\caption{Perturbative results (up to 1st order in gCS coupling $\lambda$) for $D=7$ perturbative black holes solutions in theory \refb{lhecs} (reducing to Myers-Perry (MP) black holes for $\lambda = 0$). $\bar{A}$ is the area of the perturbed horizon (see \refb{A1}). MP expressions for mass $M$, angular momenta $J_i$ and entropy $S$ are given in equations (\ref{MPmass}), (\ref{MPangm}) and (\ref{MPent}), respectively. Exact results in $\lambda$ are marked.}
\end{table}

\subsubsection{gCS terms and interior of black holes}  
\label{sssec:genJ}

Here we pause for the moment to address an interesting issue raised in \cite{Solodukhin:2005ah} on the basis of 3-dimensional analysis, which can be put as a question ``Does gravitational Chern-Simons terms see the interior of black holes?".  We shall argue here that in $D>3$ the answer is negative, and that the apparently positive answer in $D=3$ is probably a coincidence.    

Let us first state the issue. It is known that in $D=3$ the Hilbert-Einstein action supplemented with a negative cosmological constant term leads to the BTZ solutions \cite{Banados:1992wn} describing stationary rotating black holes. The difference with our problem, aside from the number of dimensions, is the presence of the negative cosmological constant term $\Lambda = - 1/\ell^2$ (which is necessary in $D=3$ if we want to have black hole solutions at all) implying that BTZ solutions are asymptotically AdS. Including a gCS Lagrangian term in $D=3$ does not affect stationary rotating black hole solutions (they are still BTZ) but does change the entropy, which can be written in the form \cite{Solodukhin:2005ah} 
\begin{equation} \label{entD3}
S = \frac{A_H}{4 \, G_N} - \mathrm{sign}(j) \frac{\beta}{\ell} \frac{A_-}{4\, G_N} \;\;, \qquad
 \beta \equiv \, 32\pi G_N \lambda
\end{equation}
where $A_-$ is the area of the inner horizon, and $j$ is angular momentum parameter. The second term comes from the gCS entropy term and we see that it depends only on a geometrical property (proper area) of the inner horizon. In \cite{Solodukhin:2005ah} it was speculated that this may not be coincidental but indicates that a gCS term may see interior of the black hole.

We investigate here the same assertion in $D>3$. In this case, as we do not known analytic solutions of  (\ref{eomD7})-(\ref{cD7}) in nontrivial cases in which contribution of the gCS entropy term is nonvanishing, we must turn to perturbation analysis in $\lambda$ around MP black hole. Using such expansion in the entropy formula (\ref{entD7}) one gets
\begin{equation} \label{entpert}
S[\bar{g}] = \frac{A_H[\bar{g}]}{4 \, G_N} + \lambda \, S_{\mathrm{gCS}}[\bar{g}]
 = \frac{A_H[\bar{g}]}{4 \, G_N} + \lambda \, S_{\mathrm{gCS}}[g_{\mathrm{MP}}] + O(\lambda^2)
\end{equation}
If $S_{\mathrm{gCS}}[\bar{g}]$ is some function of intrinsic geometric quantities connected to the inner horizon of the solution $\bar{g}$ (like, e.g., area of the inner horizon $A_-[\bar{g}]$), then 
$S_{\mathrm{gCS}}[g_{\mathrm{MP}}]$ should give the same for the MP metric $g_{\mathrm{MP}}$. 

We have already calculated this in $D=7$ and the result is presented in Eq. (\ref{entgCSMP}). We have not found any interpretation of this result in terms of geometric quantities linked to the inner horizon, or more generally, in terms of some other simple geometrical properties interior to event horizon $r_H$. This conclusion does not change if we generalize to (A)dS black holes (by introducing a cosmological constant 
$\Lambda$ in Lagrangian $\LL_0$), at least not for generic values of $\Lambda$.\footnote{This follows simply from the fact that the limit $\Lambda \to 0$ is well-defined and smooth in $D>3$, so it leads to our asymptotically flat results and corresponding conclusions.} 
 
Why and how the area of the inner horizon appears in $D=3$ in (\ref{entD3})? For our argument it is enough to restrict our attention to the more symmetric case in which all angular momenta 
are equal, which for MP black holes in $D = 2m+1$ dimensions ($m$ is an odd integer) requires $a_i = a$, $i=1,\ldots,m$. Let us assume that formula (\ref{sCSa0}) generalizes to 
\begin{equation} \label{egCSDg}
S_{\mathrm{gCS}}[g^{(0)}] = c_m \left( \frac{a}{r_H} \right)^{\!m}
\end{equation}
where $c_m$ are some constants. It is true in $D=3$ ($m=1$) because for BTZ black hole metric one has 
$a = (r_H r_-)/\ell$, where $r_- = A_-/(2\pi)$ is the radius of the inner horizon, so from (\ref{egCSDg}) Eq.\ (\ref{entD3}) follows. However, this ``mechanism'' is not possible in $D>3$, because generally
\begin{equation}
\prod_{i=1}^m |a_i| =  \prod_{i=1}^m |r_i^2|^{1/2}
\end{equation}
in the asymptotically flat case ($\Lambda = 0$), and
\begin{equation}
\prod_{i=1}^m |a_i| = \frac{1}{\ell} \prod_{i=1}^{m+1} |r_i^2|^{1/2}
\end{equation}
in the asymptotically (A)dS case ($\Lambda = \pm 1/\ell^2$), where $r_i^2$ are a complete set of 
roots of the horizon-defining polynomial  equation (Eq. (\ref{rh0eq}) in $\Lambda=0$ case). Only in 
$D=3$ one has $a^m = r_H r_-$, so that after dividing by $r_H$ one is left with $r_-$ alone in 
(\ref{egCSDg}). Other roots, aside $r_H$ and $r_-$, are not defining other inner horizons and are, as far as we know, deplete of any direct geometrical meaning. We now see that the fact that in $D=3$ one has 
$S_{\mathrm{gCS}} \propto A_-$ is probably just a coincidental consequence of the more fundamental
relation (\ref{egCSDg}).

\vspace{10pt}

\section{Perturbative calculations in $D=7$: case $a_i = a$}
\label{sec:pertD7}

\medskip

\subsection{Is perturbative expansion in $\lambda$ viable?}
\label{ssec:pertOK}

Searching for exact solutions to the equations of motion (\ref{eomD7})
\begin{equation} \label{eom2}
R_{\nu\sigma}[\bar{g}] = 16\pi G_N \lambda \, C_{\nu\sigma}[\bar{g}] \;\;,
\end{equation}
where $G_{\nu\sigma}$ is the Einstein tensor and $C_{\nu\sigma}$ the contribution of gCS Lagrangian term (\ref{gCSeom}), is probably futile. So we would like to turn to a perturbative analysis. But, of course, we have to be 
sure that a perturbative expansion in the gCS coupling $\lambda$ makes sense at all. Due to    topological reasons 
it was argued in the literature \cite{Witten:2007kt,Lu:2010sj,BCDPSp2} that only for special discrete (``quantized'') values of $\lambda$, defined through some ``quantization condition'' of the form 
\begin{equation} \label{quantCS}
\lambda_n = n \,\lambda_1 \;\;, \qquad\qquad n \in \mathrm{Z} \;\;,
\end{equation}
can one give unambiguous meaning to a gCS term in the action.\footnote{For $D=3$ it was argued in \cite{Witten:2007kt}, for $D=7$ in \cite{Lu:2010sj}, and for general case in \cite{BCDPSp2}. The argument is based on a standard application of path-integral quantization to gravity.}
The value of the constant $\lambda_1$ depends on what is exactly the space of allowed configurations.
Taken at face value, this quantization may invalidate perturbation theory in $\lambda$. 

We would like to argue that even if (\ref{quantCS}) is correct\footnote{One way to counter (\ref{quantCS}) 
is by noting that the argument used in obtaining (\ref{quantCS}) is quantum mechanical, and assumes 
that ``naive'' path integral formulation of gravity in which one integrates over metrics (or connections and vielbeins) is meaningful in nonperturbative regime. This is normally a standard quantization prescription, but gravity is hardly ``normal'' theory, especially in $D>3$ where general relativity cannot be put in the 
form of the gauge theory. Indeed, we know basically  nothing for sure about quantum gravity, so 
a skeptical view on the correctness of the quantization of gCS coupling constant is not unmotivated.}, perturbation theory in $\lambda$ can be made meaningful. One can achieve this by scaling additional parameters
of the theory, which for the stationary black holes are $G_N$, $\mu$ and $a_i$. As in this case there are 
two independent dimensionless parameters, there are several ways one can do this. We present two
possibilities:\footnote{For sake of clarity, we restrict ourself here to the case where all 
parameters $a_i$ are equal, $a_i = a$.}
\begin{itemize}
\item[(a)]
We take as two independent dimensionless parameters $c_{\lambda N} \equiv \lambda G_N / \mu^{5/4}$
and $(a/\mu^{1/4})$, and take $c_{\lambda N} \ll 1$ by making the scaling parameter 
$G_N / \mu^{5/4}$ sufficiently small while keeping $a$ and $\mu$ fixed. It is obvious that an expansion in 
$\lambda$ can be trivially written as an expansion in $c_{\lambda N}$. This is the well-known scenario when one takes Planck length $l_{\mathrm{Pl}} = G_N^{1/5}$ to be much smaller then physical scales in the problem.
\item[(b)]
We define a dimensionless parameter $c_{\lambda a} \equiv \lambda \, G_N a / \mu^{3/2}$, 
and take $c_{\lambda a} \ll 1$ by making the scaling parameter $(a / \mu^{1/4})$ sufficiently small while keeping $G_N / \mu^{5/4}$ fixed. This is meaningful because, as we show below, one can 
write the expansion in $\lambda$ as an expansion in $c_{\lambda a}$ with good convergence properties for small $a/\mu^{1/4}$. 
\end{itemize}
We are interested here in the case (b). Let us first discuss two subtleties. In both cases, (a) and (b), we can formally treat the expansion in $\lambda$ independently of the expansions of other quantities which are small in the relevant scaling parameters ($G_N / \mu^{5/4}$ and $a / \mu^{1/4}$, respectively). This is because one can make the effective coupling $c_{\lambda_n} \ll 1$ for arbitrarily high $n$ in quantization law (\ref{quantCS}), by making the relevant scaling parameter sufficiently small. However, for specifically chosen $\lambda_n$, at the end of calculation one should group all the terms with the same powers of the small scaling parameters ($G_N / \mu^{5/4}$ and $a / \mu^{1/4}$, respectively). 

We would like to argue that the claim in (b) is sound. We start from the equations of motion 
(\ref{eom2}) and consider a perturbative solution in $\lambda$ around Myers-Perry metric (\ref{mpbh}). It is obvious that a perturbative expansion for the metric can be written in the form
\begin{equation} \label{pertsol}
\bar{g}_{\nu\sigma} = \sum_{k=0}^\infty c_{\lambda N}^k \, g^{(k)}_{\nu\sigma}
\end{equation}
where $c_{\lambda N} = \lambda \, G_N / \mu^{5/4}$, $g^{(0)}_{\nu\sigma}$ is MP black hole solution with all parameters $a_i$ equal, $a_i = a$, and $g^{(k)}_{\nu\sigma}$ depend on $\mu$ and $a$ (but not on 
$\lambda$ and $G_N$). We assume that $c_{\lambda N}$ is small enough so that expansion 
(\ref{pertsol}) is convergent. If $\lambda$ is quantized, and so assumes finite value from the set 
(\ref{quantCS}), one can make $c_{\lambda N}$ small as we like by appropriately tuning Newton's constant $G_N$.

Now we want to show that in the perturbative expansion every power of $\lambda$ is accompanied by a factor of $a$. Following formally a standard procedure we insert (\ref{pertsol}) in (\ref{eom2}) and collect terms with the same power of $\lambda$. It is important to note that $g^{(0)}_{\nu\sigma}$ is analytic in 
$a$ around $a=0$, as are all operators obtained by expanding both sides in (\ref{eom2}). This allows us to make Taylor expansions in $a$. In the first order one gets (we show this
explicitly in Sec. \ref{ssec:pertD7a})
\begin{equation}
G'[g^{(0)}] \cdot g^{(1)} = C[g^{(0)}]
\end{equation}
where, for the sake of simplicity, we use an abstract notation (the symbol $G'[g^{(0)}]$ is in fact a linear differential operator acting on $g^{(1)}_{\nu\sigma}$). The key point is that right hand side (i.e.,the gCS term) generates an extra 
factor of $a^2$ (in view of the $a$-dependence), in such a way that every component in 
$g^{(1)}_{\nu\sigma}$ has an extra factor of $a$ compared with $g^{(0)}_{\nu\sigma}$. This means that if we
make the redefinition $g^{(1)}_{\rho\sigma} \equiv a^2 h^{(1)}_{\nu\sigma}$ the expansion in (\ref{pertsol}) 
becomes\footnote{In fact, as we show in Sec. \ref{ssec:pertD7a}, $g^{(1)}_{t\phi_i}$ contains a multiplicative factor of $a^2$, while all other components of $g^{(1)}_{\nu\sigma}$ have a multiplicative factor $a^3$.}
\begin{equation} \label{pert1sol}
\bar{g}_{\nu\sigma} = g^{(0)}_{\nu\sigma} + (c_{\lambda N} a^2) h^{(1)}_{\nu\sigma} +
 \sum_{k=2}^\infty c_{\lambda N}^k \, g^{(k)}_{\nu\sigma}
\end{equation}

At the second order we obtain a (differential) equation
\begin{equation} \label{1oreq}
G'[g^{(0)}] \cdot g^{(2)} = a^2 \, C'[g^{(0)}] \cdot h^{(1)} - a^4 \, G''[g^{(0)}] \cdot h^{(1)} \cdot h^{(1)}  
\end{equation}
It can be shown that $C'[g^{(0)}] \cdot h^{(1)} \propto a^2$. It then follows from (\ref{1oreq}) that 
$g^{(2)}_{\nu\sigma}$ has (at least) an extra multiplicative factor $a^4$ compared with $g^{(0)}_{\nu\sigma}$. 
Defining $g^{(2)}_{\nu\sigma} \equiv a^4 h^{(2)}_{\nu\sigma}$ and using this in (\ref{pert1sol}) we get
\begin{equation} \label{pert2sol}
\bar{g}_{\nu\sigma} = g^{(0)}_{\nu\sigma} + (c_{\lambda N} a^2) h^{(1)}_{\nu\sigma} + 
(c_{\lambda N} a^2)^2 h^{(2)}_{\nu\sigma} + \sum_{k=3}^\infty c_{\lambda N}^k \, g^{(k)}_{\nu\sigma}
\end{equation}
Repeating this procedure we finally get
\begin{equation} \label{pertsolinf}
\bar{g}_{\nu\sigma} = \sum_{k=0}^\infty (c_{\lambda N} a^2)^k h^{(k)}_{\nu\sigma}
\end{equation}
where $h^{(0)}_{\nu\sigma} \equiv g^{(0)}_{\nu\sigma}$ and all $h^{(k)}_{\nu\sigma}$ are analytic in 
$a$ around $a=0$. We now see that our perturbative expansion is an effective expansion in $(\lambda a^2)$.

We can write (\ref{pertsolinf}) in the following form
\begin{equation} \label{psolinfga}
\bar{g}_{\nu\sigma} = \sum_{k=0}^\infty (c_{\lambda a})^k \, \tilde{h}^{(k)}_{\nu\sigma} \;\;, \qquad \qquad
\tilde{h}^{(k)}_{\nu\sigma} = (\mu^{1/4} a)^k  h^{(k)}_{\nu\sigma}
\end{equation}
where $c_{\lambda a} = \lambda \, G_N a / \mu^{3/2}$ is a dimensionless parameter. What is interesting in this new parametrization is that $\tilde{h}^{(k)}_{\nu\sigma}$, beside being analytic in $a$, also satisfies
\begin{equation}
\lim_{a \to 0} \tilde{h}^{(k)}_{\nu\sigma} = 0
\end{equation}
We now see that (\ref{psolinfga}) is expansion in $(\lambda a)$ with the coefficients which become very small when $a/\mu^{1/4}$ is small, improving the convergence of the expansion in that regime. Comparing (\ref{psolinfga}) with the expansion (\ref{pertsol}), we conclude that (\ref{psolinfga}) can be made sensible even for $\lambda$ and $G_N/\mu^{5/4}$ finite, if we take $a$ small enough. This is exactly our claim in (b).

\vspace{5pt}

\subsection{Perturbative expansion in $\lambda$: Equations of motion}
\label{ssec:pertD7a}

Our aim is to find perturbative stationary rotating asymptotically flat black hole solutions in $D=7$ in a theory with Lagrangian (\ref{lhecs}) to first-order in gCS coupling $\lambda$. For simplicity we specialize to the case when all angular momenta $J_i$, $i=1,2,3$, are equal. We perturb around MP black holes which are parametrized by two numbers ($\mu$, $a$), because in this case
$a_i = a$, $i=1,2,3$.  As we discussed in Sec. \ref{sec:d7bh}, this case is rich enough to expect all relevant quantities to be perturbed by the gCS terms.

As in (\ref{pertsol}) we search for the perturbative solution 
\begin{equation} \label{pertsol1}
\bar{g}_{\nu\sigma} = g^{(0)}_{\nu\sigma} + \alpha \, g^{(1)}_{\nu\sigma} + O(\alpha^2) \;,
\end{equation}
where for convenience we defined
\begin{equation} \label{alpha}
\alpha \equiv 16\pi \, G_N \lambda \;\;.
\end{equation}
Putting (\ref{pertsol1}) in EOM (\ref{eom2}) and using gauge condition
\begin{equation} \label{gauge}
g^{(0)\nu\rho} g^{(1)}_{\nu\rho} = 0 \;\;, \qquad\qquad \nabla^{\nu} g^{(1)}_{\nu\rho} = 0,
\end{equation}
one obtains
\begin{equation} \label{eom1st}
-\frac{1}{2} \nabla^{\beta} \nabla_{\beta} \, g^{(1)}_{\nu\sigma}
 + R^{\beta}{}_{\nu\sigma}{}^{\rho} g^{(1)}_{\beta \rho} = C_{\nu\sigma}[g^{(0)}]
\end{equation}
In (\ref{gauge}) and (\ref{eom1st}) covariant derivative $\nabla_\nu$, Riemann tensor 
$R_{\beta\nu\sigma\rho}$ and $C_{\nu\sigma}$ are constructed from the unperturbed metric 
$g^{(0)}_{\nu\sigma}$, which is also used for raising and lowering indices. By solving (\ref{eom1st}) one obtains the first-order correction to metric $g^{(1)}_{\nu\sigma}$.

In our case $g^{(0)}_{\nu\sigma}$ is the MP black hole metric with all angular momenta equal, i.e., 
\begin{equation} \label{Jequal}
a_i = a \;\;, \qquad\qquad i=1,2,3 \;\;.
\end{equation} 
From (\ref{mpbh}), (\ref{fpidef}) and (\ref{mucon}) one gets
\begin{eqnarray}
ds_{(0)}^2 &\equiv& g^{(0)}_{\nu\sigma} dx^\nu dx^\sigma
\nonumber \\
 &=& -dt^2 + \frac{\mu \, r^2}{\Pi \, F} \left( dt - a \sum_{i=1}^3 \mu_i^2 \, d\phi_i \right)^{\! 2} \!
 + \frac{\Pi \, F}{\Pi - \mu \, r^2} dr^2 + (r^2 + a^2) \sum_{i=1}^3 (d\mu_i^2 + \mu_i^2 d\phi_i^2)
\label{mpbhJeq}
\end{eqnarray}
where now
\begin{equation} \label{fpidefJeq}
F = F(r) = 1 - \frac{a^2}{r^2 + a^2} = \frac{r^2}{r^2 + a^2} \; , \qquad \qquad
\Pi = \Pi(r) = (r^2 + a^2)^3
\end{equation}
Condition (\ref{Jequal}) substantially simplifies the MP metric. In fact, it can be shown that
the dependence on the coordinates $\vec{\mu}$ is completely fixed by the enhanced symmetries induced 
by (\ref{Jequal}). We use this to write $g^{(1)}_{\nu\sigma}$ in the following form
\begin{eqnarray}
ds_{(1)}^2 &\equiv& g^{(1)}_{\nu\sigma} dx^\nu dx^\sigma
\nonumber \\
 &=& f_{t}(r) \left(\mu - (a^2 + r^2)^2 \right) dt^2 + f_{r}(r) \frac{r^2 (a^2 + r^2)^2}{\Pi - \mu r^2} dr^2 + 
  h(r) (a^2+r^2) ( d\mu_i^2 +\mu_i^2 d\phi_i^2 )
 \nonumber \\ 
&& - f_{t\phi}(r) \frac{a \, \mu}{(a^2+r^2)^2} \mu_i^2 dt \, d\phi_i 
 + f_{\phi}(r) \frac{a^2\mu}{(a^2 + r^2)^2} \mu_i^2 \mu_j^2 d\phi_i d\phi_j
\label{g1Jeq}
\end{eqnarray}
where $f_t$, $f_r$, $h$, $f_{t\phi}$, and $f_\phi$ are five unknown functions of the coordinate $r$ alone, to be
found by solving the equations of motion. We see that in the special case (\ref{Jequal}), due to the enhancement of symmetry, the problem generally (i.e., not only in perturbation theory) boils down to solving a system of \emph{ordinary} differential equations, which is of immense help.

Writing (\ref{mpbhJeq}) and (\ref{g1Jeq}) in the gauge conditions (\ref{gauge}) imposes two constraints on
unknown functions, which we use to express $f_t(r)$ and $f_{t\phi}(r)$ in terms of the remaining three
functions. Using this in the EOM (\ref{eom1st}) one gets the system of second-order differential equations for the remaining unknown functions $f_r(r)$, $h(r)$, and $f_\phi(r)$. As these equations are rather long they are presented in Appendix \ref{app:tech:DE}.

\vspace{5pt}

\subsection{Solving at lowest order in $a$}
\label{ssec:solalaw}

Equations (\ref{freq})-(\ref{heq}) still appear nasty enough to be solved exactly, so we turn to slowly 
rotating black holes, i.e., $a/\mu^{1/4} \ll 1$.\footnote{A similar double perturbative perturbative expansion was performed in \cite{Yunes:2009hc} for the case of perturbation of Einstein gravity with massless scalar field in $D=4$ by a mixed Chern-Simons Lagrangian term. In contrast to our case, in this theory the lowest-order correction does not capture changes in the horizon properties like area and temperature.} In this regime, solutions of (\ref{freq})-(\ref{fpeq}), with proper asymptotic behavior to describe asymptotically flat black holes, are given by 
\begin{eqnarray}
f_r(r) &=& \frac{432}{5} \frac{a^3 \mu^3}{r^{16} \left(r^4-\mu \right)} + O(a^5)
\nonumber \\
f_\phi(r) &=& -1296 \, \frac{a \, \mu^2}{r^{14}} - \frac{5 \, r^6}{a^2 \mu} h(r) + O(a^3)
\nonumber \\
h(r) &=& \frac{2592}{5} \frac{a^3}{\mu^2}\, \tilde{h}(r^4/\mu) + O(a^5) \;\;,
\label{sola}
\end{eqnarray}
where the function $\tilde{h}(u)$ is given by
\begin{eqnarray}
\tilde{h}(u) &=& - Q_{1/2}(2u-1) \int_1^u \frac{dx}{x^5} P_{1/2}(2x-1)
\nonumber \\
 && + P_{1/2}(2u-1) \left( \int_\infty^u \frac{dx}{x^5} Q_{1/2}(2x-1)
  - i \frac{\pi}{2} \int_1^\infty \frac{dx}{x^5} P_{1/2}(2x-1) \right)
\label{hhdef}
\end{eqnarray}
and $P_\nu$ and $Q_\nu$ are standard Legendre functions. Using (\ref{sola}) and (\ref{hhdef}) in (\ref{g1tpi}) we obtain
$g^{(1)}_{\mu\nu}$ at the lowest order in $a$
\begin{subequations}
\allowdisplaybreaks
\begin{align}
g^{(1)}_{tt} &= - \frac{6048}{5} \frac{a^3 \mu^3}{r^{20}} + O(a^5)
 \\
g^{(1)}_{t\phi_i} &= - \frac{72}{5} \frac{a^2 \mu^3 (43 \, r^4 - 45\, \mu)}{r^{18}(r^4 - \mu)} \mu_i^2
 + O(a^4)
 \\
g^{(1)}_{rr} &= \frac{432}{5} \frac{a^3 \mu^3}{r^{12} \left(r^4-\mu \right)^2} + O(a^5)
 \\
g^{(1)}_{\mu_1\mu_1} &= \frac{2592}{5} \frac{a^3}{\mu^2} r^2 \tilde{h}(r^4/\mu)
 \frac{1-\mu_2^2}{1-\mu_1^2-\mu_2^2} + O(a^5)
 \\
g^{(1)}_{\mu_1\mu_2} &= \frac{2592}{5} \frac{a^3}{\mu^2} r^2 \tilde{h}(r^4/\mu)
 \frac{\mu_1 \mu_2}{1-\mu_1^2-\mu_2^2} + O(a^5)
 \\
g^{(1)}_{\mu_2\mu_2} &= \frac{2592}{5} \frac{a^3}{\mu^2} r^2 \tilde{h}(r^4/\mu)
 \frac{1-\mu_1^2}{1-\mu_1^2-\mu_2^2} + O(a^5)
 \\
g^{(1)}_{\phi_i\phi_j} &= - 1296 \frac{a^3 \mu^3}{r^{18}}
 \left[ 1 + 2 \frac{r^{20}}{\mu^5} \tilde{h}(r^4/\mu) \right] \mu_i^2 \mu_j^2
 + \delta_{ij} \frac{2592}{5} \frac{a^3}{\mu^2} r^2 \, \tilde{h}(r^4/\mu) \, \mu_i^2 + O(a^5)
\end{align}
\label{g1a1o}%
\end{subequations}
where $i,j = 1,2,3$ and $\mu_3^2 = 1 - \mu_1^2 - \mu_2^2$. Note that the ``ugly'' part containing 
$\tilde{h}$ cancels in $g^{(1)}_{tt}$ and $g^{(1)}_{t\phi_i}$ in the lowest order in $a$.

Let us check that our perturbed solution still describes an asymptotically flat black hole.  We will do this by
checking the behavior of the perturbed metric in two limits - asymptotic infinity and near-horizon. For this
we need the corresponding behavior of the function $\tilde{h}(u)$ which we defined in (\ref{hhdef}).

The asymptotic behaviour of the function $\tilde{h}(u)$ in the $u \to \infty$ limit is of the form
\begin{equation} \label{hhasymp}
\tilde{h}(u) = C u^{-3/2}
 + O(u^{-5/2}) \;\;.
\end{equation}
where the constant $C$ is
\begin{equation} \label{cval}
C = - \frac{\pi}{16} \int_1^\infty \frac{dx}{x^5} P_{1/2}(2x-1) \approx -0.0593 \ldots
\end{equation}
This means that $\tilde{h}(r^4/\mu) \propto 1/r^6$, so the asymptotic behavior of (\ref{g1a1o}) at
the limit $r \to \infty$ in the lowest order in $a$ is 
\begin{equation} \label{falloff}
g^{(1)}_{tt} \sim O(r^{-20}) \;, \quad  g^{(1)}_{t\phi_i} \sim O(r^{-18}) \;, \quad
 g^{(1)}_{rr} \sim O(r^{-20}) \;, \quad g^{(1)}_{\mu_i\mu_j} \sim O(r^{-4}) \;, \quad
 g^{(1)}_{\phi_i\phi_j} \sim O(r^{-4}) \;.
\end{equation}
We see explicitly that the perturbed solution is still asymptotically flat and that the fall-off conditions 
(\ref{falloff}) guarantee that the metric perturbation (\ref{g1a1o}) does not change the relations between 
asymptotic quantities (energy and angular momentum) and black hole parameters ($\mu$ and $a$).
However, we should ask what happens in higher orders in the perturbation parameter $a/\mu^{1/4}$. To 
answer this we have performed a detailed analysis by perturbatively solving eqs.(\ref{freq})-(\ref{fpeq}) 
in the regime $r \gg \mu^{1/4}$, using (\ref{g1a1o}) as starting point, to all relevant orders in 
$u = r^4/\mu$ and $a/\mu^{1/4}$. We have found that $g^{(1)}_{\mu\nu}$ has the following asymptotic 
behavior at $r \to \infty$
\begin{equation} \label{fallofftrue}
g^{(1)}_{tt} \sim \frac{a^5}{r^{16}} \;, \quad  g^{(1)}_{t\phi_i} \sim \frac{a^4}{r^{10}} \;, \quad
 g^{(1)}_{rr} \sim \frac{a^5}{r^{12}} \;, \quad g^{(1)}_{\mu_i\mu_j} \sim \frac{a^3}{r^4} \;, \quad
 g^{(1)}_{\phi_i\phi_j} \sim \frac{a^3}{r^4} \;.
\end{equation}  
We see that after including \emph{all} orders of $a/\mu^{1/4}$ in $g^{(1)}_{\mu\nu}$, asymptotics have changed, but not in a significant way - the conclusion is that Myers-Perry relations 
(\ref{MPmass}) and (\ref{MPangm}) are still valid up to first-order in $\lambda$.\footnote{Eq. 
(\ref{Qpseudo}) implicitly says that the gCS term could possibly contribute to $M$ and $J$ only if some of the components of the metric perturbation had asymptotic behavior 
$\delta g_{tt} \sim r^{-4}$, $\delta g_{t\phi} \sim r^{-4}$, $\delta g_{rr} \sim r^{-4}$ and 
$\delta g_{ij} \sim r^{-2}$. But, we see from (\ref{fallofftrue}) that all the components have faster fall-off, so there is no contribution to $M$ or $J$.}

In the limit $r \to \mu^{1/4}$ (i.e., $u \to 1$), the function $\tilde{h}$ has the following expansion
\begin{equation} \label{thexpu1}
\tilde{h}(u) = \tilde{h}(1) + \frac{1}{4} (77 + 23 \, \tilde{h}(1)) (u-1)
 + \frac{5}{64} (847 + 173 \, \tilde{h}(1)) (u-1)^2 + O(u^3)
\end{equation}
where
\begin{equation} \label{thu1}
\tilde{h}(1) = - \int_1^\infty \frac{dx}{x^5} \left( Q_{1/2}(2x-1) + i \frac{\pi}{2} P_{1/2}(2x-1) \right)
 \approx - 0.15336 \ldots
\end{equation}
This implies that the metric perturbation (\ref{g1a1o}) has the expected behavior for
the black hole in the vicinity of the horizon, which, at the zeroth-order in $a$, is located at 
$r = \mu^{1/4}$. We shall see that part of the expansion (\ref{thexpu1}) proportional to the ``ugly" 
constant $\tilde{h}(1)$ does not contribute to near-horizon quantities (event horizon and ergosurface\footnote{Generically, the ergosurface is not in the near-horizon region. However, as we are doing a
perturbative calculation in $a$, for $|a|/\mu^{1/4} \ll 1$ the ergosurface is perturbatively close to the 
horizon.} properties), as it cancels in the calculations.

Now we are ready to calculate corrections to various black hole parameters. Below we present
the main results while technical details of the calculations can be found in Appendix \ref{app:tech:pertq}.

\subsubsection{Event horizon}
\label{sssec:evhor}

We can find the location of the event horizon in standard fashion from
\begin{equation} \label{rHeq}
\bar{g}^{rr}(\bar{r}_H) = 0
\end{equation}
From (\ref{pertsol1}), (\ref{mpbhJeq}) and (\ref{g1a1o}) follows
\begin{equation}
\bar{g}^{rr}(r) = \frac{(r^2 + a^2)^3 - r^2 \mu}{r^2 (r^2 + a^2)^2}
 - \alpha \left( \frac{432}{5} \frac{a^3 \mu^3}{r^{20}} + O(a^5) \right)
 + \sum_{k=2}^\infty \alpha^k \, O(a^{2k})
\end{equation}
which, plugged in (\ref{rHeq}), gives
\begin{equation} \label{rhoriz}
\bar{r}_H = r_{H0} + \alpha \left( \frac{108}{5} \frac{a^3}{\mu^{7/4}} + O(a^5) \right)
 + \sum_{k=2}^\infty \alpha^k \, O(a^{2k})
\end{equation}
where $r_{H0}$ is the horizon radius for a MP black hole with $a_i =a$. Taking into account
the possibility that the gCS coupling constant $\lambda$ (which by (\ref{alpha}) implies the same for $\alpha$) 
is quantized, we must eventually view formally the double expansion in (\ref{rhoriz}) (over $\alpha$ and $a$) 
as a single expansion (over $a$).
\footnote{We explained this in detail in Sec. \ref{ssec:pertOK}.} 
The final result is
\begin{equation} \label{rHapert}
\bar{r}_H = \mu^{1/4} - \frac{3}{4} \frac{a^2}{\mu^{1/4}} + \frac{108}{5} \frac{\alpha \, a^3}{\mu^{7/4}}
 + O(a^4)
\end{equation}

We note that the same result for the event horizon is obtained from an analysis of circular orbits, which leads to the horizon condition
\begin{equation}
\left( \sum_i g_{t\phi_i} \right)^{\!\!2} - g_{tt} \sum_{i,j} g_{\phi_i \phi_j} = 0
\end{equation}
The details can be found in Appendix \ref{app:tech:pertq}.

The location of the horizon is a coordinate dependent result, so by itself the result 
(\ref{rhoriz})-(\ref{rHapert}) does not say much.\footnote{Note that (\ref{rhoriz})-(\ref{rHapert}) naively suggest that for $a>0$ the gCS term tends to``enlarge" the horizon (at lowest order of perturbation around $a=0$), but calculating the horizon area (\ref{areaH}) shows that it actually tends to ``shrink" it.} We 
have to calculate proper, coordinate independent, quantities connected with event horizon. One such obvious is the proper area of the horizon, which we also need to find the black hole entropy. In Appendix \ref{app:tech:area} we show that it is given by
\begin{equation} \label{areaH}
\bar{A}_H = A_H^{(0)}
 - \alpha \left( 540 \pi^3 \frac{a^3}{\mu^{3/4}} + O(a^5) \right)
+ \sum_{k=2}^\infty \alpha^k \, O(a^{2k})
\end{equation}
where the first term on the right side is the horizon area of the Myers-Perry black hole
\begin{equation} \label{MP7DA}
A_H^{(0)} = \pi^3 \mu \, r_{H0}
\end{equation}
By expanding $r_{H0}$ (horizon radius of the MP black hole) in $a$ we obtain
\begin{equation} \label{aHapert}
\bar{A}_H = \pi^3 \left( \mu^{5/4} - \frac{3}{4} a^2 \mu^{3/4}
 - 540 \, \alpha \frac{a^3}{\mu^{3/4}} \right) + O(a^4)
\end{equation}
Now we see that the gCS Lagrangian term induces a real change on geometry of black hole solutions.

\subsubsection{Ergosurface}
\label{sssec:ergos}

The location of the ergosurface is obtained from the infinite red-shift condition
\begin{equation} \label{rerdef}
\bar{g}_{tt}(\bar{r}_e) = 0 \;\;.
\end{equation}
From (\ref{pertsol1}), (\ref{mpbhJeq}) and (\ref{g1a1o})  
\begin{equation}
\bar{g}_{tt}(r) = -1 + \frac{\mu}{(r^2 + a^2)^2}
 - \alpha \left( \frac{6048}{5} \frac{a^3 \mu^3}{r^{20}} + O(a^5) \right)
 + \sum_{k=2}^\infty \alpha^k \, O(a^{2k})
\end{equation}
follows. 
By inserting this in (\ref{rerdef}) we obtain that the ergosurface is defined by the condition $r=\bar{r}_e$,
where 
\begin{equation} \label{rergo}
\bar{r}_e = \sqrt{\mu^{1/2} - a^2} - \alpha \left( \frac{1512}{5} \frac{a^3}{\mu^{7/4}} + O(a^5) \right)
 + \sum_{k=2}^\infty \alpha^k \, O(a^{2k})
\end{equation}
By expanding the first term (which is the ergosurface radius of the MP black hole) and collecting powers of $a$ we obtain
\begin{equation}
\bar{r}_e = \mu^{1/4} - \frac{1}{2} \frac{a^2}{\mu^{1/4}} - \frac{1512}{5} \frac{\alpha \, a^3}{\mu^{7/4}}
 + O(a^4)
\end{equation}

\subsubsection{Angular velocity}
\label{sssec:angvel}

If we write the horizon generating null Killing vector $\bar\chi$ as
\begin{equation} \label{killgen}
\bar\chi = \frac{\partial}{\partial t} + \bar\Omega_H \sum_i \frac{\partial}{\partial \phi_i}
\end{equation}
then $\bar\Omega_H$ is the angular velocity of the horizon. We can obtain it from the null-condition
on the horizon
\begin{equation}
\bar{\chi}^2(\bar{r}_H) \equiv \left. \bar\chi^\mu \bar\chi^\nu \bar g_{\mu\nu}\right|_{r=\bar r_H} = 0
\end{equation}
From (\ref{killgen}) and the form of the metric it follows
\begin{equation} \label{avform}
\bar{\Omega}_H
 = - \left. \frac{\sum_i \bar g_{t\phi_i} }{\sum_{i,j} \bar g_{\phi_i\phi_j}} \right|_{r=\bar r_H}
\end{equation}
Putting (\ref{pertsol1}), (\ref{mpbhJeq}), (\ref{g1a1o}), and (\ref{rhoriz}) in (\ref{avform}) we obtain
\begin{equation} \label{avsol}
\bar{\Omega}_H = \frac{a}{r_{H0}^2 + a^2}
 - \alpha \left( 648 \frac{a^2}{\mu^{2}} + O(a^4) \right)
 + \sum_{k=2}^\infty \alpha^k \, O(a^{2k}) 
\end{equation}
By expanding the first term (which is $\Omega_H$ of the MP black hole) and collecting powers of 
$a$ we obtain
\begin{equation}
\bar{\Omega}_H = \frac{a}{\sqrt{\mu}} - 648 \, \alpha \frac{a^2}{\mu^{2}}
 + \frac{1}{2} \frac{a^3}{\mu} + O(a^4)
\end{equation}

\subsubsection{Surface gravity and black hole temperature}
\label{sssec:temper}

The surface gravity $\bar{\kappa}$ is defined by
\begin{equation}
\bar\chi^\mu \bar\nabla_\mu \bar\chi^\nu = \bar\kappa \bar\chi^\nu  \qquad
 \textrm{on the horizon $r = \bar r_H$}
\end{equation}
Using (\ref{killgen}), (\ref{avsol}), (\ref{pertsol1}), (\ref{mpbhJeq}), (\ref{g1a1o}), and (\ref{rhoriz})
we obtain
\begin{equation}
\bar\kappa = \frac{3 r_{H0}}{r_{H0}^2 + a^2}-\frac{1}{r_{H0}} 
+ \alpha \left( 1944 \frac{a^3}{\mu^{9/4}} + O(a^5) \right)
 + \sum_{k=2}^\infty \alpha^k \, O(a^{2k})
\end{equation}
By expanding the first term (which is $\kappa$ of the MP black hole) and collecting powers of 
$a$ we obtain
\begin{equation}
\bar\kappa = \frac{2}{\mu^{1/4}} - \frac{3}{2} \frac{a^2}{\mu^{3/4}}
+ 1944 \, \alpha  \frac{a^3}{\mu^{9/4}} + O(a^4)
\end{equation}
The black hole temperature $\bar{T}_H$ is obtained from surface gravity via
\begin{equation}
\bar{T}_H = \frac{\bar\kappa}{2\pi}
\end{equation}

\subsubsection{Black hole entropy}
\label{sssec:entropy}

As discussed in Sec. \ref{sec:eom} a black hole entropy in our case is given by
\begin{equation} \label{Stot}
\bar{S}_{\mathrm{bh}} = \bar{S}_{\mathrm{BH}} + \lambda \bar{S}_{\mathrm{gCS}}
\end{equation}
where $S_{\mathrm{BH}}$ is the Bekenstein-Hawking entropy proportional to the proper horizon area
$A_H$
\begin{equation}
\bar{S}_{\mathrm{BH}} = \frac{\bar{A}_H}{4 \, G_N}
\end{equation}
and $S_{\mathrm{gCS}}$ is the contribution induced by Lagrangian gCS term given by 
\cite{Tachikawa:2006sz,Bonora:2011gz}
\begin{equation}
S_{\mathrm{gCS}} =  16\pi \int_\mathcal{B} \Gam_N \R_N^2
\end{equation}
By using (\ref{pertsol1}), (\ref{mpbhJeq}), (\ref{g1a1o}), (\ref{rhoriz}) and (\ref{alpha}) we obtain
\begin{equation} \label{SBH}
\bar{S}_{\mathrm{BH}} = \frac{A_H^{(0)}}{4 \, G_N}
 - \lambda \left( 2160 \pi^4 \frac{a^3}{\mu^{3/4}} + O(a^5) \right)
+ \sum_{k=2}^\infty \alpha^k \, O(a^{2k})
\end{equation}
where the first term is the entropy of the Myers-Perry black hole. In the same way we obtain
\begin{equation} \label{SgCS}
\bar{S}_{\mathrm{gCS}} = 3456 \, \pi^4  \left( \frac{a}{r_{H0}} \right)^{3}
 + \sum_{k=2}^\infty \alpha^k \, O(a^{2k})
\end{equation}
Interestingly the simple result (\ref{SgCS}) is $a$-exact in lowest order in $\lambda$. Plugging 
(\ref{SBH}) and (\ref{SgCS}) into (\ref{Stot}) gives us the black hole entropy
\begin{equation}
\bar{S}_{\mathrm{bh}} = \frac{A_H^{(0)}}{4 \, G_N}
 + \lambda \left( (6 \pi)^4 \frac{a^3}{\mu^{3/4}} + O(a^5) \right)
 + \sum_{k=2}^\infty \alpha^k \, O(a^{2k})
\end{equation}
or, written purely as expansion in $a$,
\begin{equation}
\bar{S}_{\mathrm{bh}} = \frac{\pi^3}{4 \, G_N} \left( \mu^{5/4} - \frac{3}{4} a^2 \mu^{3/4} \right)
 + (6 \pi)^4 \lambda \frac{a^3}{\mu^{3/4}} + O(a^4)
\end{equation}

\subsubsection{Mass, angular momentum and the 1st law of BH thermodynamics}
\label{sssec:1stlaw}

We have mentioned above that if, to calculate asymptotic charges, we use the method based on the energy-momentum pseudotensor, then the asymptotic fall-off (\ref{falloff}) of our perturbative solution guarantees that mass $M$ and angular momentum $J$ do not receive gCS corrections to the lowest orders in the perturbative expansion (first order in $\lambda$ and $a^3$ ($a^2$) for $M$ ($J$)). In mathematical terms, the result is
\begin{equation} \label{MJanon}
M = M_0 + \lambda\, O(a^5) + O(\lambda^2) \;, \qquad  
J = J_0 + \lambda\, O(a^4) + O(\lambda^2) 
\end{equation}
Strictly speaking, to be fully consistent with the first law of black hole thermodynamics one would need to calculate energy and angular momentum using Wald's procedure adapted to theories with gCS terms 
\cite{Iyer:1994ys,Tachikawa:2006sz,Bonora:2011gz}. As we do not have a proof that Wald's method would give the same result as the pseudotensor method we have used, we need to check that our results are in agreement with the 1st law of BH thermodynamics. The first law in our case has the form
\begin{equation} \label{1stlaw}
\delta M = T\, \delta S + 3\, \Omega\, \delta J
\end{equation}
By using $\delta = d\mu\, \partial/\partial\mu + da\, \partial/\partial a$, our perturbative results for $T$, $S$ and $\Omega$, together with the first law for the MP black hole, from (\ref{1stlaw}) we obtain
\begin{eqnarray} \label{cond1}
\frac{\partial (M - M_0)}{\partial\mu}
 - \frac{3\,a}{\sqrt{\mu}} \, \frac{\partial (J - J_0)}{\partial\mu}
 = \lambda\, O(a^5) + O(\lambda^2)
\\
\frac{\partial (M - M_0)}{\partial a}
 - \frac{3\,a}{\sqrt{\mu}} \, \frac{\partial (J - J_0)}{\partial a}
 = \lambda\, O(a^4) + O(\lambda^2)
\label{cond2}
\end{eqnarray}
We note that the terms proportional to $\lambda a^3$ nontrivially cancel on the right hand side of 
(\ref{cond1}), and likewise the terms proportional to $\lambda a^2$ on the right hand side of (\ref{cond2}).

We see that our result (\ref{MJanon}) is fully consistent with (\ref{cond1})-(\ref{cond2}), i.e., with the first law of black hole thermodynamics. Even more, it can be shown that (\ref{cond1})-(\ref{cond2}) and symmetry properties necessarily imply (\ref{MJanon}).

We have argued, based on the asymptotic analysis of Appendix \ref{app:asymp}, that the full first-order gCS correction to the metric has, at $r\to\infty$, the asymptotic behavior given in (\ref{fallofftrue}). From this the stronger result for $M$ and $J$
\begin{equation} \label{MJnon}
M = M_0 + O(\lambda^2) \;, \qquad  
J = J_0 + O(\lambda^2) 
\end{equation}
follows.
Though we are quite confident that this result is correct, we cannot check its full consistency with the first law of BH thermodynamics, because we cannot calculate other thermodynamical parameters ($T$, 
$\Omega$ and $S$) to the desired order (that is to the first order in $\lambda$ and to all orders in $a$).

\vspace{10pt}

\section{Conclusion}
\label{sec:concl}

\medskip

We have investigated in some detail the influence of adding a purely gravitational Chern-Simons Lagrangian term in the action of some diffeomorphism covariant theory of gravity, on asymptotically flat stationary rotating black hole solutions and corresponding black hole entropy. We have shown that the structure of the Chern-Simons term, characterized by its parity violating properties, does not have any effect when two or more angular momenta vanish. Perturbative arguments indicate that, instead, in 
cases when at most one angular momentum is zero, the influence of gravitational Chern-Simons terms is nontrivial, both on the solutions and the entropy.

In an attempt to find black hole solutions we have specialized to what seems to be the simplest nontrivial case, i.e.
Einstein gravity supplemented with a gravitational Chern-Simons Lagrangian term in $D=7$, and black holes with all angular momenta equal. We have calculated the first-order correction of the gravitational Chern-Simons entropy term and argued that it does not correspond to any geometric property of the interior of black hole (like the inner horizon surface area), contrary to the conjecture made in 
\cite{Solodukhin:2005ah} which was based on the analysis of rotating AdS black holes in $D=3$. Due to the complexity of the equations of motion, 
we have not been able to find exact analytic solutions. We have turned to a double perturbative expansion, in Chern-Simons coupling constant and angular momentum, and constructed the first-order correction to Myers-Perry solution. We have explicitly calculated corrections to horizon area, ergoregion and black hole entropy, all of which are nonvanishing. A perturbative analysis shows that the influence of the gravitational Chern-Simons Lagrangian term is completely nontrivial: it changes the type of metric - the perturbed metric does not seem to fall into the Kerr-Schild class with a flat seed metric. This is unfortunate because the Kerr-Schild ansatz was the crucial tool used for constructing Kerr and Myers-Perry solutions. It remains to be seen whether our perturbative results can suggest some new ansatz which could be used in analytic constructions. 

An obvious extension would be to include a cosmological constant and consider asymptotically (A)dS solutions. In Einstein gravity exact analytic solutions of this type were obtained in \cite{Gibbons:2004js}, so one could naively expect that extension of our treatment to this case should be straightforward. However this is not the case - introduction of cosmological constants seriously complicates the calculations. This interesting problem is currently under investigation.

\vspace{20pt}

{\bf Acknowledgments}

\bigskip

\noindent
One of us (L.B.) would like to thank the Theoretical Physics Department, University of Zagreb, for hospitality and financial support during his visits there. M.C., P.D.P., S.P.\ and I.S.\ would like to thank SISSA for hospitality and financial support during visits there and would also like to acknowledge support by the Croatian Ministry of Science, Education and Sport under the contract no.~119-0982930-1016. The work of L.B. was supported in part by the MIUR-PRIN contract 2009-KHZKRX. Visits have been financially supported in part by the University of Zagreb Development Fund.

\vspace{10pt}

\section*{Appendix}
\appendix

\medskip

\section{Some technical details}
\label{app:tech}

\subsection{Differential equations for metric correction}
\label{app:tech:DE}

Writing (\ref{mpbhJeq}) and (\ref{g1Jeq}) in the gauge conditions (\ref{gauge}) imposes two constraints on
unknown functions, which we use to express $f_t(r)$ and $f_{t\phi}(r)$ in terms of the remaining three
functions. The result is that the metric correction (\ref{g1Jeq}) can be written as
\begin{subequations}
\allowdisplaybreaks
\begin{align}
g^{(1)}_{tt} &= \frac{1}{3r(r^2 + a^2)^3} \left\{ 2r \left[\mu \, a^2 f_{\phi}(r)
 + \left( 4(r^2 + a^2)^3 - \mu (2 r^2 + a^2) \right) f_r(r) \right. \right.
\nonumber \\
 & \left. \left.  + \left( 5(r^2 + a^2)^3 - 2 a^2 \mu \right) h(r) \right]
 + (r^2 + a^2) \left( (r^2 + a^2)^3 - r^2 \mu \right) f'_r(r) \right\}
 \\
g^{(1)}_{t\phi_i} &= \frac{\mu_i^2}{6a\mu r(r^2 + a^2)^3} \left\{ \mu a^2 r \left( 
 \mu (3 r^2 + 5 a^2) - (r^2 + a^2)^3 \right) f_{\phi}(r) \right. 
 \nonumber \\
 & + r \left( 5(r^2 + a^2)^6 - \mu (r^2 - 6 a^2) (r^2 + a^2)^3 - 2 \mu^2 a^2 (2 r^2 + a^2) \right) f_r(r)
 \nonumber \\
 & - r \left( 5(r^2 + a^2)^6 - 3 \mu (5 r^2 + 3 a^2) (r^2 + a^2)^3 + 4 \mu^2 a^4 \right) h(r)
 \nonumber \\
 & \left. + (r^2 + a^2) \left( (r^2 + a^2)^6 - \mu (r^2 - a^2) (r^2 + a^2)^3  - r^2 a^2 \mu^2 \right) f'_r(r)
  \right\}
 \\
g^{(1)}_{rr} &= \frac{r^2 (r^2+a^2)^2}{(r^2+a^2)^3-r^2 \mu} f_r(r)
 \\
g^{(1)}_{\mu_1\mu_1} &= \frac{(r^2+a^2) (1-\mu_2^2)}{1-\mu_1^2-\mu_2^2} h(r)
 \\
g^{(1)}_{\mu_1\mu_2} &= \frac{(r^2+a^2) \mu_1 \mu_2}{1-\mu_1^2-\mu_2^2} h(r)
 \\
g^{(1)}_{\mu_2\mu_2} &= \frac{(r^2+a^2) (1-\mu_1^2)}{1-\mu_1^2-\mu_2^2} h(r)
 \\
g^{(1)}_{\phi_i\phi_j} &=
\frac{a^2 \mu \, \mu_i^2\mu_j^2}{(r^2+a^2)^2} f_\phi(r) + \delta_{ij} (r^2+a^2) \mu_i^2 \, h(r)
\end{align}
\label{g1tpi}%
\end{subequations}
Inserting (\ref{mpbhJeq}) and (\ref{g1tpi}) in the EOM (\ref{eom1st}) we obtain the following system 
of differential equations for the remaining unknown functions $f_r(r)$, $h(r)$, and $f_\phi(r)$
\begin{eqnarray}
f''_r(r) &=& \frac{\mu \, r^2 (7 r^2 + 3 a^2) - (15 r^2 - a^2)(r^2 + a^2)^3}{r (r^2 + a^2)
 \left[ (r^2 + a^2)^3 - \mu \, r^2 \right]} f'_r(r)
 - \frac{8 r^2 \left( 5 (r^2 + a^2)^3 - a^2 \mu \right)}{(r^2 + a^2)^2
 \left[ (r^2 + a^2)^3 - \mu \, r^2 \right]} f_r(r)
 \nonumber \\
 && + \frac{8 a^2 \mu \, r^2}{(r^2 + a^2)^2 \left[ (r^2 + a^2)^3 - \mu \, r^2 \right]} f_\phi(r)
 + \frac{8 r^2 \left( 5 (r^2 + a^2)^3 - 2 a^2 \mu \right)}{(r^2 + a^2)^2
 \left[ (r^2 + a^2)^3 - \mu \, r^2 \right]} h(r)
 \nonumber \\[1mm]
 && + \frac{3456 \, a^3 \mu^3 r^2 (7 r^2 - a^2)}{(r^2 + a^2)^{10}
 \left[ (r^2 + a^2)^3 - \mu \, r^2 \right]}
\label{freq} \\[3mm]
h''(r) &=& \frac{2r}{r^2 + a^2} f'_r(r) + \frac{4 r^2 \left( 2 (r^2 + a^2)^3 - a^2 \mu \right)}{(r^2 + a^2)^2
 \left[ (r^2 + a^2)^3 - \mu \, r^2 \right]} f_r(r)
 \nonumber \\[1mm]
 && - \frac{4 a^2 \mu \, r^2}{(r^2 + a^2)^2 \left[ (r^2 + a^2)^3 - \mu \, r^2 \right]} f_\phi(r)
 - \frac{(5 r^2 - a^2)(r^2 + a^2)^2 - \mu \, r^2}{r \left[ (r^2 + a^2)^3 - \mu \, r^2 \right]} h'(r)
 \nonumber \\[1mm]
 && - \frac{8 r^2 \left( (r^2 + a^2)^3 - a^2 \mu \right)}{(r^2 + a^2)^2
 \left[ (r^2 + a^2)^3 - \mu \, r^2 \right]} h(r)
 + \frac{1728 \, a^3 \mu^3 r^2}{(r^2 + a^2)^9 \left[ (r^2 + a^2)^3 - \mu \, r^2 \right]}
\label{fpeq}
\end{eqnarray}
\begin{eqnarray}
f''_\phi(r) &=& - \frac{2r \left( 5 (r^2 + a^2)^3 + 4 a^2 \mu  \right)}{a^2 \mu (r^2 + a^2)} f'_r(r)
 - \frac{4 r^2 \left( 10 (r^2 + a^2)^6 + 3 a^2 \mu (r^2 + a^2)^3 - 4 a^4 \mu^2 \right)}{a^2 \mu
 (r^2 + a^2)^2 \left[ (r^2 + a^2)^3 - \mu \, r^2 \right]} f_r(r)
 \nonumber \\[1mm]
 && + \frac{5 r^2 + a^2}{r (r^2 + a^2)} f'_\phi(r)
 + \frac{4 r^2 \left( 5 (r^2 + a^2)^3 + 4 a^2 \mu \right)}{(r^2 + a^2)^2
 \left[ (r^2 + a^2)^3 - \mu \, r^2 \right]} f_\phi(r)
\nonumber \\[1mm]
 && - \frac{2 r \left( 5 (r^2 + a^2)^6 - 3 \mu (5 r^2 + a^2) (r^2 + a^2)^3 + 4 a^4 \mu^2 \right)}{a^2 \mu
 (r^2 + a^2) \left[ (r^2 + a^2)^3 - \mu \, r^2 \right]} h'(r) 
 \nonumber \\[1mm]
 && + \frac{8 r^2 \left( 5 (r^2 + a^2)^6 - a^2 \mu (r^2 + a^2)^3 - 4 a^4 \mu^2 \right)}{a^2 \mu
 (r^2 + a^2)^2 \left[ (r^2 + a^2)^3 - \mu \, r^2 \right]} h(r)
  \nonumber \\[1mm]
 && - \frac{1728 \, a \, \mu^2 r^2 \left( (215 r^2 - 57 a^2) (r^2 + a^2)^3
 - 2 \mu (105 r^4 - 33 a^2 r^2 - 2 a^4) \right)}{(r^2 + a^2)^{10}
 \left[ (r^2 + a^2)^3 - \mu \, r^2 \right]}
\label{heq}
\end{eqnarray}

\subsection{Perturbed horizon, ergosphere, angular velocity, surface gravity}
\label{app:tech:pertq}

We are interested in finding the position of the horizon (see e.g.\ \cite{Frolov:1998} page 63,  \cite{Padmanabhan:2010zzb} page 373) by looking at the massless test particle 
in the circular motion around the black hole\footnote{Our conventions are essentially the same as in 
\cite{Wald:1984rg,Iyer:1994ys}, except for the definition of the binormal to the horizon $\epsilon_{\mu\nu}$ which has the opposite sign.}
$$u = \frac{\partial}{\partial t} + \omega_i \frac{\partial}{\partial \phi_i} $$
Because of the symmetry induced by $a_i=a$ it suffices to restrict ourselves to the case $\omega_i = \Omega$ where $\Omega$ is found by solving $u^2=0$. Now, we use the fact that the horizon is the surface where there is only one solution for $\Omega$. Therefore the horizon position $r_H$ is found from
\be \label{eqrhor}
\left(\sum_i g_{t\phi_i}\right)^2 - g_{tt} \sum_{i,j} g_{\phi_i\phi_j} = 0
\ee
We apply the same prescription to the perturbed metric $\gu+\dg$. To first order we have:
\be
&&\delta r \frac{\partial}{\partial r} \left\lbrace \left(\sum_i \gu_{t\phi_i}\right)^2 - \gu_{tt} \sum_{i,j} \gu_{\phi_i\phi_j} \right\rbrace
+ \\&& \quad {} +
2\left(\sum_i \gu_{t\phi_i}\right) 
\left( \sum_i \dg_{t\phi_i}\right) - \gu_{tt} \sum_{i,j} \dg_{\phi_i\phi_j}
- \dg_{tt} \sum_{i,j} \gu_{\phi_i\phi_j} = 0 \0
\ee
Plugging in the explicit expressions we obtain:
\be
\delta r = \frac{(a^2 + r^2) (\Pi - r^2 \mu) (f_r(r) + 4h(r))}{2r((\Pi + (r^2-a^2)\mu))}\bigg|_{r=r_H}
\ee
We note that the new horizon is located on $r = \textrm{const}$, i.e.\ $\delta r$ is not a function of $\mu_i$.
We assume that $h(r)$ and $(\Pi - r^2 \mu) f_r(r)$ are regular at the horizon $r=r_H$ (the explicit solution \refb{sola} justifies this). 
For convenience we define $f_{r2}(r)$ to be
\be \label{deffr2}
f_{r2}(r) = (\Pi(r) - r^2 \mu)f_r(r)
\ee
which is regular and nonvanishing at the horizon. 
We obtain
\be \label{dr}
\delta r = \frac{(a^2 + r^2)  f_{r2}(r) }{2\mu r(2r^2-a^2)}\bigg|_{r=r_H}
\ee
We can check this result by using the formula $\gp^{rr} = 0$ which is valid when the horizon is at 
$r = \textrm{const}$ (see e.g.\ \cite{Poisson} page 190-191). We get
\be \delta r = \frac{\gu^{rr}f_r(r)}{\partial_r \gu^{rr} }   \bigg|_{r=r_H}
= \frac{(a^2 + r^2) f_{r2}(r)}{2\mu r(2r^2 - a^2)} \bigg|_{r=r_H}
\ee
which is the same as \refb{dr}.

Next, we give expressions for perturbed angular velocity, surface gravity and radius of the ergosurface
in terms of the ansatz functions
\be 
\bar{\Omega}
=\frac{a}{a^2+r_H^2} &-&
\left(
-\frac{\left(2 a^4+13 a^2 r_H^2+2 r_H^4\right) f_{r2}(r_H)}
{6 a \left(a^2-2 r_H^2\right) \left(a^2+r_H^2\right)^2 \mu }
-
\frac{f_{r2}'(r_H)}
{6 a r_H \mu }
\right. \0\\
&& +\left.\frac{ a(a^2-2r_H^2)(a^2+r_H^2) f_\phi(r_H)}
{6 r_H^2 \mu }
\right. \0\\
&& +\left.\frac{\left(a^2-2 r_H^2\right) \left(a^2+r_H^2\right) \left(4 a^2+5 r_H^2\right) h(r_H)}
{6 a r_H^2 \mu }
\right) + O(\lambdaexppar)^2 \0
\ee
\be
\bar\kappa =
\frac{3 r_H}{a^2+r_H^2}-\frac{1}{r_H}
&+&  \left(
\frac{\left(-2 a^4-13 a^2 r_H^2-2 r_H^4\right) f_{r2}(r_H)}
{2 r_H \left(a^2-2 r_H^2\right) \left(a^2+r_H^2\right)^2 \mu }
-
\frac{f_{r2}'(r_H)}{2 r_H^2 \mu }+
\right. \0\\
&& \quad +\left.
\frac{\left(a^4-2 a^2 r_H^2\right) f_\phi (r_H)}
{2 r_H \left(a^2+r_H^2\right)^2}
+
\right. \0\\
&& \quad +\left.
\frac{\left(a^2-2 r_H^2\right) \left(4 a^2+5 r_H^2\right) h(r_H)}
{2 r_H \left(a^2+r_H^2\right)^2}
\right) + O(\lambdaexppar)^2 \0
\ee
\be
\bar r_e  = \sqrt{\sqrt{\mu}-a^2} + \frac{(a^2+r^2) f_{r2}'(r)}{12 \mu  r^2}+
\frac{a^2 f_\phi(r)}{6 r}+\left(
\frac{5 r}{6}-
\frac{a^2}{3 r}\right) h(r)+
\frac{f_{r2}(r)}{6 \mu  r}
\ee

\subsection{Area of the perturbed horizon: $A_{\Hp}[\gp]$}
\label{app:tech:area}
 
The area is
\be \label{A1}
A_{\Hp}[\gp] = \int_{\Hp} \sqrt{q(\Hp,\gp)} d^5 x = 
 (2\pi)^3 \int_0^1 d\mu_1 \int_0^{\sqrt{1-\mu_1^2}} d\mu_2 
\sqrt{q(\Hp,\gp)} 
\ee
where $q(H,g)_{\sigma\nu}$ is the metric induced on the horizon $H$ from the metric $g$ 
\be\label{defq}
q(H,g)_{\sigma\nu} = g_{\sigma\nu} + n_{\sigma} l_{\nu} + l_{\sigma} n_{\nu}  
\ee
where $n^\nu$ and $l^\nu$ are null normals to $H$ normalized as $n^\nu l_\nu = -1$.
We wish to express \refb{A1} in terms of unperturbed quantities. 
We denote $q_0 = q(\Hu,\gu)$. We use the formula $\delta \sqrt q = \frac{1}{2} \sqrt q q^{\sigma\nu} \delta q_{\sigma\nu}$ to relate the square roots of the determinants of the following metrics induced on the  perturbed horizon (to first order):
\be 
\sqrt{q(\Hp,\gp)} - \sqrt{q(\Hp,\gu)} 
=  \frac{1}{2} \sqrt{q_0} q_0{}^{ab} \dg_{ab} 
=  \sqrt{q_0} n^\sigma l^\nu \dg_{\sigma\nu} \label{qgg}
\ee 
where we used \refb{defq} and the gauge condition \refb{gauge}. 
Here,  $n^\nu$ and $l^\nu$ denote a pair of null vectors that are normal to the unperturbed horizon $\Hu$ with respect to the unperturbed metric $\gu$.
Next, we express the difference $\sqrt{q(\Hp,\gu)} - \sqrt{q(\Hu,\gu)}$ to first order using $\delta r$:
\be \label{qhh}
\sqrt{q(\Hp,\gu)} - \sqrt{q_0} =  \delta r \partial_r \sqrt{q_0} 
\ee
Denoting by $A_{H(r)}[\gu]$ the area of the surface $H(r)$ (defined by $r=\textrm{const}$, $t=\textrm{const}$) measured by the unperturbed metric, 
i.e.\ $A_{H(r)}[\gu] = \int_{H(r)} \sqrt{q(H(r),\gu)} d^5 x$, for the 7-dimensional Myers-Perry geometry we obtain
\be \label{AHR}
A_{H(r)}[\gu] = \pi ^3 \left(a^2+r^2\right) \sqrt{\left(a^2+r^2\right)^3+a^2 \mu }
\ee
When $r=r_H$, this reduces to \refb{MP7DA}. Using \refb{qgg}, \refb{qhh} and \refb{AHR}, the area \refb{A1} becomes 
\be
A_{\Hp}[\gp] = A_{\Hu}[\gu]  + \delta r \partial_r  A_{H(r)}[\gu] 
+    A_{\Hu}[\gu]   n^\sigma l^\nu \dg_{\sigma\nu}  + O(\lambdaexppar^2) 
\ee
The second term on the right hand side is
\be
\delta r \partial_r A_{H(r)}[\gu]\bigg|_{r=r_H} = \delta r A_{\Hu}[\gu] \left( \frac{3 r_H^{5/3}}{\mu^{2/3}}+\frac{2 r_H^{1/3}}{\mu^{1/3}} \right)
\ee
To calculate $n^\sigma l^\nu \dg_{\sigma\nu}$ we use
\be
n_\nu = \frac{1}{\sqrt{2}} \left( \frac{dt_\nu}{\sqrt{-g^{tt}}} + \frac{dr_\nu}{\sqrt{g^{rr}}} \right) \qqd
 l_\nu = \frac{1}{\sqrt{2}} \left( \frac{dt_\nu}{\sqrt{-g^{tt}}} - \frac{dr_\nu}{\sqrt{g^{rr}}} \right)
\ee
and the ansatz for $\dg$ \refb{g1tpi}
\be
n^\sigma l^\nu \dg_{\sigma\nu} \bigg|_{r=r_H} = \frac{1}{2} \left(  \left(1-\frac{r_H^{4/3}}{\mu^{1/3}}\right)f_\phi (r_H)+  \left(4+\frac{r_H^{4/3}}{\mu^{1/3}}\right)h(r_H)\right)
\ee
Now we can write the area in terms of the ansatz functions 
\be
A_{\Hp}[\gp] &=&  A_{\Hu}[\gu] \left( 1 + 
 \frac{f_{r2}(r_H)}{2r_H^2 - a^2}
\left( \frac{3 r_H^{4/3}}{2 \mu^{4/3}}+\frac{1}{\mu} \right)
\right.\0\\ && + \left.
\frac{1}{2} \left(  \left(1-\frac{r_H^{4/3}}{\mu^{1/3}}\right)f_\phi(r_H) +  \left(4+\frac{r_H^{4/3}}{\mu^{1/3}}\right)h(r_H)\right)
\right) + O(\lambdaexppar^2) 
\ee
where $f_{r2}(r)$ is defined in \refb{deffr2}. Finally plugging in the explicit solutions \refb{sola}-\refb{hhdef} gives \refb{areaH} and \refb{aHapert}.

\subsection{Entropy: $\Sent_{\mathrm{gCS}}[\gu]$}
\label{app:tech:S1g0}

Now we give the details of the calculation of \refb{entgCSMP}. Here $\gu_{\rho\sigma}$ denotes 
the unperturbed 7-dimensional Myers-Perry solution  \refb{mpbh}, and $\Gamma^c{}_{\rho\sigma}$ and $R_{abcd}$ are constructed from $\gu_{\rho\sigma}$. 
In $D=7$, Eq.\ \refb{entgCS} becomes
\be \label{S1}
\Sent_{\mathrm{gCS}}[\gu] 
&=& 16 \pi   \int_{\Hu} \Gam_{N} \R_{N}^2 \0\\
&=& 16 \pi   \int_{\Hu} \Gam_{N} \wedge \R_{N} \wedge \R_{N} \0\\
&=& 16 \pi  \frac{5!}{4} \int_{\Hu} 
(\Gamma_{N})_{[\sigma_1} (R_{N})_{\sigma_2 \sigma_3} (R_{N})_{\sigma_4 \sigma_5]} \0\\
&=& \frac{16 \pi}{4} \int_{\Hu} \varepsilon^{\sigma_1 \ldots \sigma_{5}}
 (\Gamma_{N})_{\sigma_1} (R_{N})_{\sigma_2 \sigma_3} (R_{N})_{\sigma_4 \sigma_5} d^{5} x
\ee
Here $\Gamma_{N\beta} = \frac{1}{2} \epsilon_{\sigma\rho} \Gamma^{\rho\sigma}{}_\beta$, 
$R_{N\beta\gamma} = \frac{1}{2} \epsilon_{\sigma\rho} R^{\rho\sigma}{}_{\beta\gamma}$ 
and $\epsilon_{\mu\nu}$ is the binormal to the horizon, normalized as $\epsilon_{\mu\nu}\epsilon^{\mu\nu} = -2$ with $\epsilon_{tr} < 0$.\footnote{This normalization is due to the definition of $\epsilon_{\rho\sigma}$ in Eq.\ (4.7) of \cite{Bonora:2011gz} as 
$\nabla_\sigma \xi^\rho \big|_\mathcal{B} = \kappa \, \epsilon^\rho_{\;\; \sigma}$.}
Also, the totally antisymmetric tensor density 
$\varepsilon$  satisfies
$\varepsilon^{3 \ldots {D}} = 1$. 
We have $-g^{tt} g^{rr}= \frac{r^2 \mu + (\Pi - r^2\mu)F}{F^2 \Pi}$, so the binormal to the horizon $r=r_H$
is $\epsilon_{tr} = -\frac{1}{\sqrt{-g^{tt} g^{rr}}} = -F(r_{H},\mu_{1},\mu_{2})$. 
Plugging $\Gamma$ and $R$ in \refb{S1} we get
\be
\Sent_{\mathrm{gCS}}[\gu] 
= -\frac{16 \pi}{4}\; (2\pi)^3 \; 32\; a_1 a_2 a_3 \;  \mu^2 r_H \; Q\;  I ,
\ee
where
\be \label{Qdef}
Q &=& \frac{1}{2} \sum_{i \ne j} (r_H^2 + a_i^2) (r_H^2 + a_j^2)
 = \sum_{i=1}^3 \frac{\mu r_H^2}{r_H^2 + a_i^2}
\0\\ &=&
a_1^2 a_2^2  + a_2^2 a_3^2 + a_1^2 a_3^2 + 2 r_H^2 ( a_1^2 + a_2^2 + a_3^2 ) + 3r_H^4
\ee
and
\be
I = \frac{1}{4} \int_0^1 d\mu_1^2 \int_0^{1-\mu_1^2} d\mu_2^2
  \frac{
 9r_H^4 + \sum_{i=1}^3 ( 2 a_i^4 \nu_i^4 + 9 r_H^2  a_i^2 \nu_i^2 )
- \sum_{i<j} a_i^2 a_j^2 ( \mu_i^2 \mu_j^2 - 5 \nu_i^2 \nu_j^2 ) }
{ (r_H^4 + a_1^2 a_2^2 \mu_3^2 + a_2^2 a_3^2 \mu_1^2 + a_1^2 a_3^2 \mu_2^2  + r_H^2 \sum_{i=1}^3 a_i^2 \nu_i^2 )^5 }
  \0
\ee
where $\nu_i^2 = 1 - \mu_i^2$, $\mu_3^2 = 1 - \mu_1^2 - \mu_2^2$.
This can be integrated, the result is $I = \frac{Q^2}{8 \mu^4 r_H^8}$. The final result for $\Sent_{\mathrm{gCS}}[\gu]$ is then
\be \label{ent10i}
\Sent_{\mathrm{gCS}}[\gu]  = 128 \, \pi^4  \frac{\mu}{r_H} a_1 a_2 a_3
 \left( \sum_{i=1}^3 \frac{1}{r_H^2 + a_i^2} \right)^3
\ee

\bigskip

\section{Perturbative calculation in the asymptotic infinity region}
\label{app:asymp}

In Section \ref{ssec:solalaw} we indicated that our perturbative solution (\ref{g1a1o}) (lowest order in 
$a$)\footnote{It is implicitly assumed that we are still restricted to the first-order correction in the gCS coupling $\lambda$.} does not necessarily give a leading order contribution at asymptotic infinity 
$r \to \infty$, in an expansion in $1/r$. This means that we still have to convince ourself that (\ref{g1a1o}) indeed corresponds to some exact solution which is asymptotically flat Minkowski with finite mass and angular momenta. The idea is to perturbatively solve the system (\ref{g1tpi}) as expansion in 
$1/r$ such that in the lowest order in $a$ this new perturbative solution is consistent with (\ref{sola}). After some work one gets the following asymptotic perturbative solution
\begin{eqnarray}
f_r(r) &=& - h_0 \frac{a^5 \sqrt{\mu}}{r^{12}} \left\{ 2 - 12 \frac{a^2}{r^2}
 + \left( 42 a^4 + \frac{19}{8} \mu \right) \frac{1}{r^4}
 + O\left( r^{-6} \right) \right\} + \frac{a^3 \mu^3}{r^{20}} \left\{ \frac{432}{5} + O\left( r^{-2} \right) \right\}
\nonumber \\
h(r) &=& \frac{h_0\, a^3}{\sqrt{\mu}\, r^6} \left\{ 1 - 3 \frac{a^2}{r^2} + \left(6 a^4 + \frac{3}{4} \mu \right)
 \frac{1}{r^4} - \left( 10 a^4 + \frac{15}{4} \mu \right) \frac{a^2}{r^6}
 + \left( 15 a^8 + \frac{45}{4} a^4 \mu + \frac{75}{128} \mu^2 \right) \frac{1}{r^8}
\right. \nonumber \\
&& \left. - \left( 21 a^8 + \frac{105}{4} a^4 \mu + \frac{573}{128} \mu^2 \right) \frac{a^2}{r^{10}}
 + O\left( r^{-{12}} \right) \right\}
 + \frac{a^3 \mu^3}{r^{20}} \left\{ \frac{5184}{385} + O\left( r^{-2} \right) \right\}
\nonumber \\
f_\phi(r) &=& - \frac{h_0 a}{\mu^{3/2}} \left\{ 5 + \frac{15}{4} \frac{\mu}{r^4}
 - \frac{11}{2} \frac{a^2 \mu}{r^6} + \left( \frac{21}{4} a^4 + \frac{375}{128} \mu \right) \frac{\mu}{r^8}
 - \left( 3 a^4 \mu + \frac{375}{32} \mu \right) \frac{a^2 \mu}{r^{10}}
 + O\left( r^{-{12}} \right) \right\}
\nonumber \\
 && - \frac{a\, \mu^2}{r^{14}} \left\{ \frac{104976}{77} + O\left( r^{-2} \right) \right\}
\label{fhfpsi}
\end{eqnarray}
Comparison with full space solution (in lowest order in $a$) (\ref{sola}) fixes an integration constant $h_0$ to be $h_0 = 2592\,C/5$, where $C$ is defined in (\ref{cval}). In (\ref{fhfpsi}) the rightmost terms in all three equations (terms which are not proportional to $h_0$) are written just to show full asymptotic structure of our solution, their fall-off is too fast to affect leading order asymptotic behavior of the metric tensor to which we now turn. 

Plugging (\ref{fhfpsi}) into (\ref{g1tpi}) we obtain the following asymptotic expansion for the first order correction in $\lambda$ of the metric tensor: 
\begin{eqnarray}
g^{(1)}_{tt} &=& - \frac{5}{4} h_0 \frac{a^5 \mu^{3/2}}{r^{16}} + O\left(r^{-18}\right)
\label{g1ttinf} \\
g^{(1)}_{t\phi_i} &=& - \frac{8}{3} h_0 \frac{a^4 \sqrt{\mu}}{r^{10}} + O\left(r^{-12}\right)
\label{g1tpinf} \\
g^{(1)}_{rr} &=& - 2 h_0 \frac{a^5 \sqrt{\mu}}{r^{12}} + O\left(r^{-12}\right)
\label{g1rrinf} \\
g^{(1)}_{\mu_i\mu_j} &=& \chi_{ij}(\vec{\mu}) \left(
 h_0 \frac{a^3}{\sqrt{\mu}\, r^4} + O\left(r^{-6}\right) \right)
\label{g1mminf} \\
g^{(1)}_{\phi_i\phi_j} &=& \delta_{ij} \mu_i^2 
 \left( \frac{h_0 a^3}{\sqrt{\mu}\, r^4} + O(r^{-6}) \right)
 + \mu_i^2 \mu_j^2 \left( - 5 \frac{h_0 a^3}{\sqrt{\mu}\, r^4} + O(r^{-6}) \right)
\label{g1ppinf}
\end{eqnarray}

\vspace{10pt}


\end{document}